\documentclass{ieeeaccess}

\usepackage{amsmath,mdwlist,amssymb,amsfonts,amsthm,mathrsfs}
\usepackage{bm,cite}
\usepackage{algorithmic}

\usepackage{epsfig} 
\usepackage{graphicx,latexsym} 
\usepackage{color,soul}
\usepackage{threeparttable}
\usepackage{tabularx,multirow}

\begin{document}
\history{Date of publication xxxx 00, 0000, date of current version xxxx 00, 0000.}
\doi{10.1109/ACCESS.2017.DOI}

\title{A CMOS SoC for Wireless Ultrasonic Power/Data Transfer and SHM Measurements on Structures}

\author{\uppercase{Xinyao Tang}\authorrefmark{1}, \IEEEmembership{Member, IEEE}, \uppercase{Soumyajit Mandal}\authorrefmark{1}, \IEEEmembership{Senior Member, IEEE}, \uppercase{and Tayfun Ozdemir}\authorrefmark{2}, \IEEEmembership{Senior Member, IEEE}}
\address[1]{Department of Electrical, Computer, and Systems Engineering, Case Western Reserve University, Cleveland, OH 44106, USA (e-mail: \{xxt81, sxm833\}@case.edu)}
\address[2]{Virtual EM, Inc., Ann Arbor, MI 33174, USA (e-mail: tayfun@virtualem.com)}
\tfootnote{This work was partially funded by an U.S. Air Force STTR Phase I Contract awarded to Virtual EM Inc. (Contract No. FA9550-17-C-0001).}

\markboth
{Tang \headeretal: CMOS SoC for Wireless SHM}
{Tang \headeretal: CMOS SoC for Wireless SHM}

\corresp{Corresponding author: Soumyajit Mandal (e-mail: soumya@alum.mit.edu).}

\begin{abstract}
This paper describes a highly-integrated CMOS system-on-chip (SoC) for active structural health monitoring (SHM). The chip integrates ultrasonic power and bidirectional half-duplex data transfer, a power management unit (PMU), and an ultrasound transceiver to enable wireless ultrasonically-coupled sensor SHM networks on structures. The PMU includes an active bias-flip rectifier with off-delay compensation, high-efficiency dual-path DC-DC converter with inductor time-sharing, and five switched-capacitor DC-DC converters to generate multi-level spectrally band-limited pulses for guided-wave SHM. The chip was fabricated in a standard 180~nm process and has a die area of $2\times 2$ mm$^{2}$. Test results show power conversion efficiency (PCE) $>85\%$ for the active rectifier, $>70$\% for the inductive DC-DC converter, and $>60$\% for the switched-capacitor DC-DC converters. Output pulses have a peak-to-sidelobe ratio (PSL) $>30$~dB and worst-case out-of-band emissions $<-30$~dB, respectively. The SoC was integrated with a low-power microcontroller and passive components to realize miniaturized (15~mm $\times$ 30~mm) wireless SHM nodes. A set of nodes was deployed on an SHM test-bed (carbon fiber reinforced polymer sheet) representing an airframe panel. Tests on this wireless network confirm both long-range ultrasound power/data transfer and the ability to detect structural damage.
\end{abstract}


\titlepgskip=-15pt
\maketitle

\section{Introduction}
\PARstart{M}{uch} of the world's critical civil infrastructure, including bridges, pipelines, and transportation, is in increasingly poor condition due to the effects of ageing and deferred maintenance~\cite{hall2016future}. Structural health monitoring (SHM) is fast becoming an important component of any integrated strategy for managing the associated risks~\cite{brownjohn2007structural,raghavan2007review,farrar2006introduction}. 

Broadly speaking, SHM sensors can be divided into two categories: \emph{passive} (sensing-only) and \emph{active} (both actuation and sensing). Scalable SHM systems of both types require small, lightweight, inexpensive, unobtrusive, and minimally invasive sensor networks~\cite{hu2014large}. Traditional wired SHM networks use individual wires for power and data transfer to each sensor node. The large number of wires required to support a large-scale wired network of this type presents installation and maintenance challenges. In addition, the weight of these wires is often unacceptable for high-value structures, such as airframes. Replacing point-to-point wired links with a single wired bus can reduce the weight of the wires, but introduces significant reliability challenges since a single-point bus failure can now disable the entire network. Using a RF-based wireless network can eliminate the wires~\cite{sazonov2004wireless}, but at the expense of significantly higher power consumption (and hence lower operating lifetime). These networking issues are particularly challenging for active SHM nodes, which provide greater measurement flexibility (since actuation waveforms can be arbitrarily selected) but also have higher power consumption. Thus, while self-powered passive SHM nodes (typically using vibration energy harvesting) have been demonstrated~\cite{lee2016ultralow,huang2011asynchronous,salehi2018damage}, similar progress on wireless active SHM nodes is lacking.

These fundamental issues can be addressed by exploiting the high efficiency of directional ultrasound links for power/data transfer, as already demonstrated for biomedical implants~\cite{seo2016wireless,charthad2015mm}. In fact, one can envision using guided acoustic waves propagating through the structure for both sensing and power/data transfer to the SHM nodes, thus enabling methods for jointly optimizing all three processes~\cite{li2018wirelessly,malzer2019combined}. We refer to such through-structure wireless networks for active SHM as being ultrasonically-coupled~\cite{shaik2020self}. 

This paper describes the design and testing of a custom CMOS IC that enables miniaturized wireless sensor nodes for ultrasonically-coupled active SHM networks. Each autonomous node relies on this IC for wireless acoustic power and data transfer, power management, and making SHM measurements. The rest of this paper is organized as follows. Wireless SHM networks are introduced in Section~\ref{sec:shm}. The design of the IC is presented in Section~\ref{sec:chip_design}. Electrical characterization results from the IC are discussed in Section~\ref{sec:electrical_results}, while Section~\ref{sec:shm_results} presents measurement results obtained from an SHM test bed. Finally, Section~\ref{sec:conclusion} concludes the paper.

\section{Introduction to Wireless SHM Networks}
\label{sec:shm}

\subsection{SHM Measurements}
Guided ultrasound waves are popular for SHM of thin-walled structures such as airframes, submarine hulls, storage tanks, and pipes. These ``Lamb waves'' can propagate relatively long distances with little loss, thus allowing a few sensors to monitor large areas of the structure using transmission (pitch-catch), reflection (pulse-echo), or passive (impact/acoustic emission detection) measurements~\cite{su2006guided,giurgiutiu2007structural}. 

\begin{figure}
    \centering
    \includegraphics[width=0.50\columnwidth]{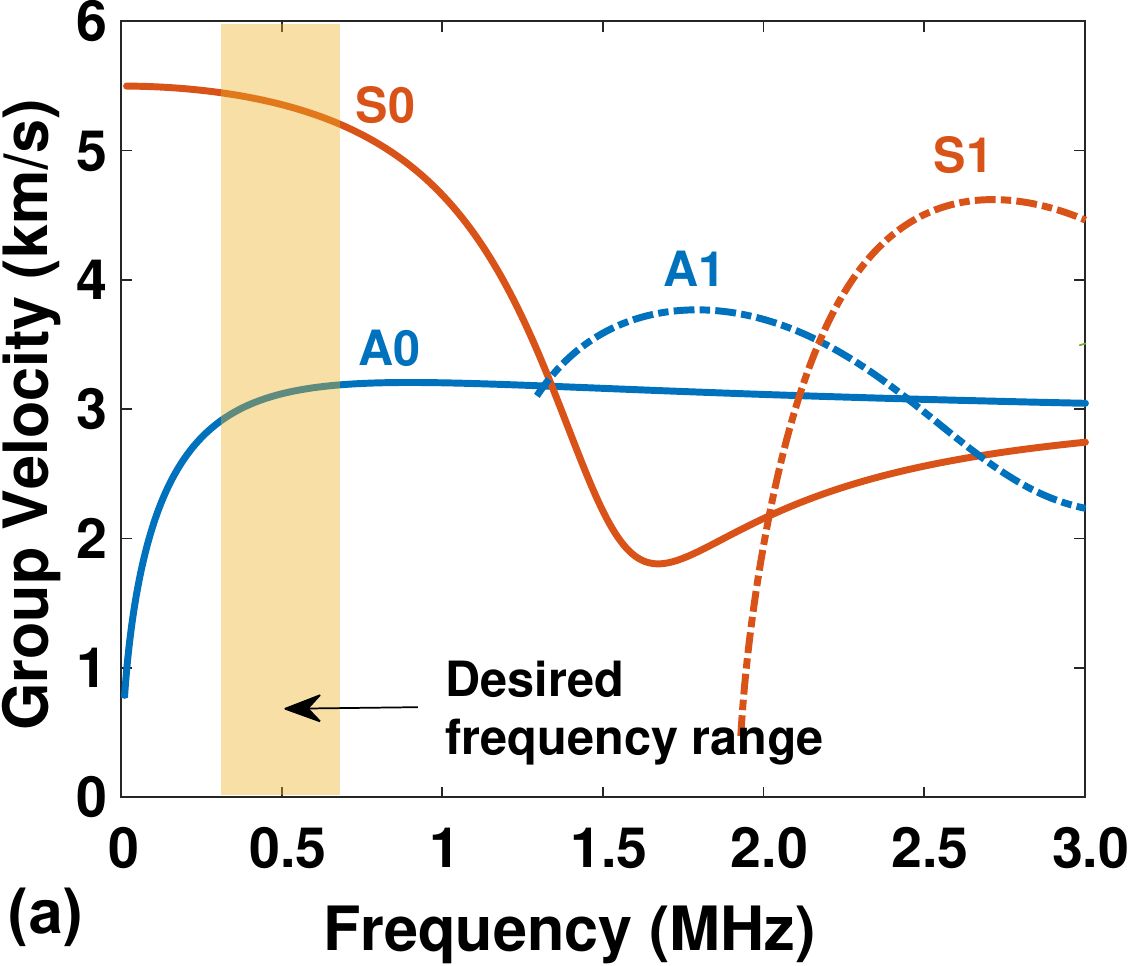}
    \includegraphics[width=0.46\columnwidth]{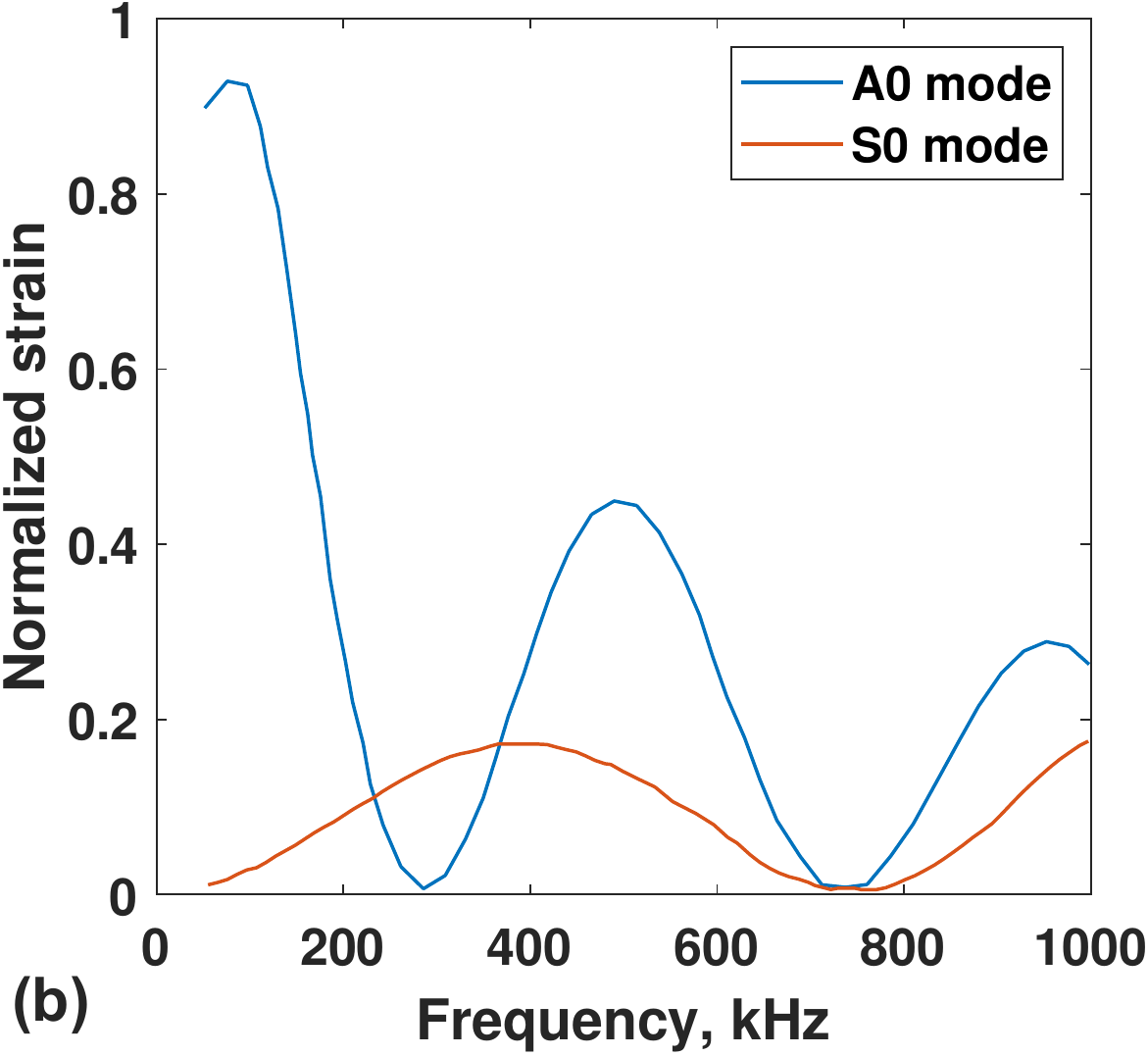}
    \caption{Active SHM using PWAS on a 1.5~mm-thick aluminum plate: (a) dispersion of Lamb waves; and (b) expected amplitudes of $S_0$ and $A_0$ modes versus frequency for a 7~mm-square PWAS (figure adapted from~\cite{giurgiutiu2003lamb}).}
    \label{fig:lamb_waves}
\end{figure}

Multiple Lamb waves modes - either symmetric (denoted $S_{i}$ where $i\geq 0$), or asymmetric (denoted $A_{i}$) - can propagate in thin-walled structures, as shown in Fig.~\ref{fig:lamb_waves}(a). Moreover, the figure shows that they are generally dispersive, i.e., have frequency-dependent group velocities. To simplify subsequent signal processing and damage detection algorithms, most applications assume that only one or two wave modes (typically $A_0$ and $S_0$) have been excited. A band-limited excitation signal can be used for this purpose, as shown in the figure. Note that the required center frequency is material- and structure-dependent: it scales as $v_s/d$ where $v_s$ is the shear wave velocity in the material and $d$ is the thickness.

Thin piezoelectric transducers, which are also known as piezoelectric wafer active sensors (PWAS), are popular for generating and receiving Lamb waves due to their low profile, broadband characteristics, and low cost~\cite{giurgiutiu2003lamb}. Their lateral (i.e., in-plane) vibrations excite Lamb wave modes in a frequency-dependent manner, as shown in Fig.~\ref{fig:lamb_waves}(b). For example, in this case, operating around 300~kHz ensures that only the $S_0$ mode is excited. Thus, the spectral properties (center frequency, bandwidth, and side-lobe levels) of SHM transmitters must be precisely controlled to obtain the best monitoring results.

Both transmission- and reflection-type SHM measurements require pulsed waveforms to obtain spatial resolution. However, simple ``on-off'' pulses with rectangular amplitude profiles are undesirable because of their poor spectral side-lobe levels (worst-case of $-13$~dB). Thus, windowed pulses should be used instead. For example, one can use cosine-sum or raised cosine window functions, which are defined as
\begin{equation}
    w[n]=a_0 + \left(1-a_0\right)\cdot\cos\left(\frac{2\pi n}{N}\right),\quad 0\leq n\leq N,
\end{equation}
where $a_0$ is a constant. The choice $a_0=25/46\approx 0.54$ results in the Hamming window, which is favored for active SHM due its low worst-case side lobe level (approximately $-41$~dB).

\subsection{Ultrasound Power Transfer on Structures}
\begin{figure}
    \centering
    \includegraphics[width=0.49\columnwidth]{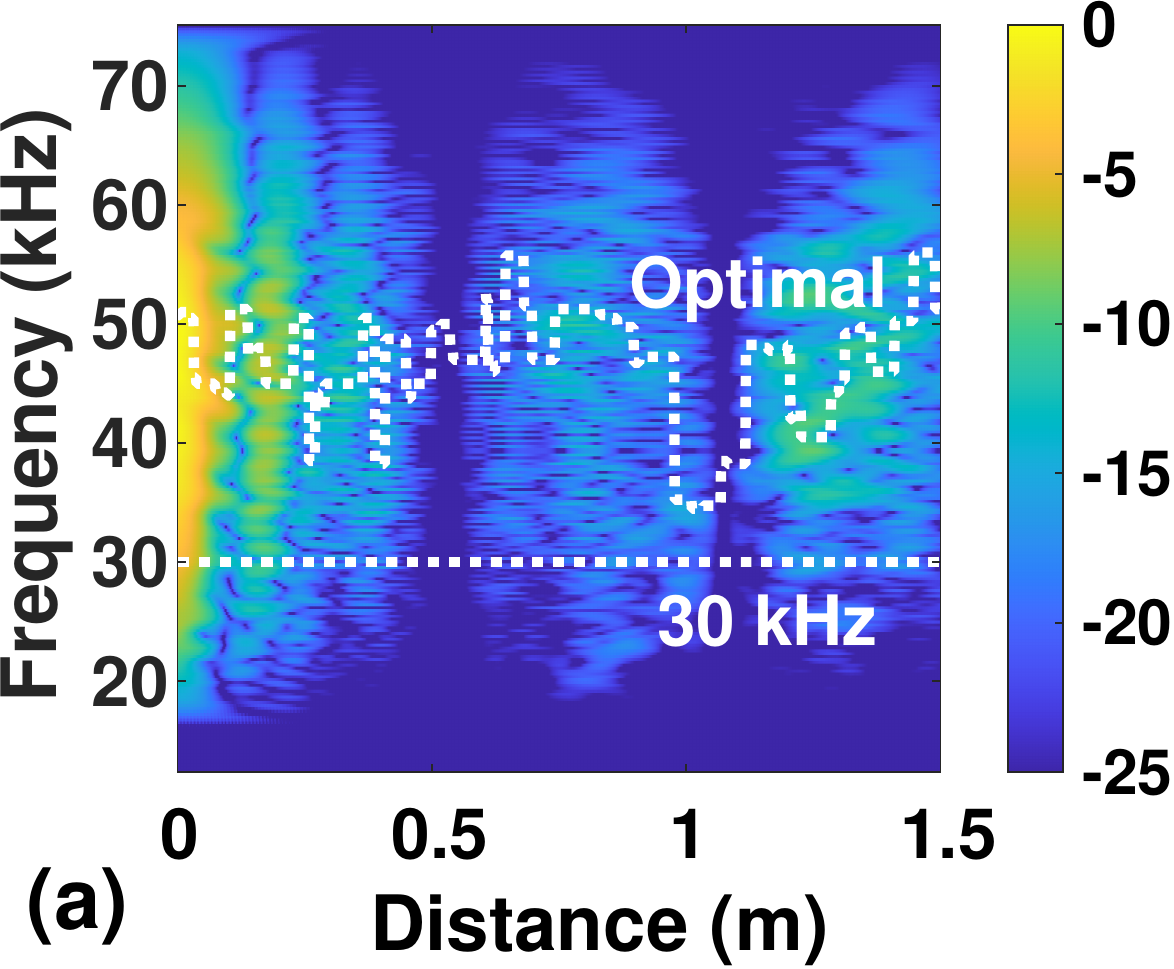}
    \includegraphics[width=0.47\columnwidth]{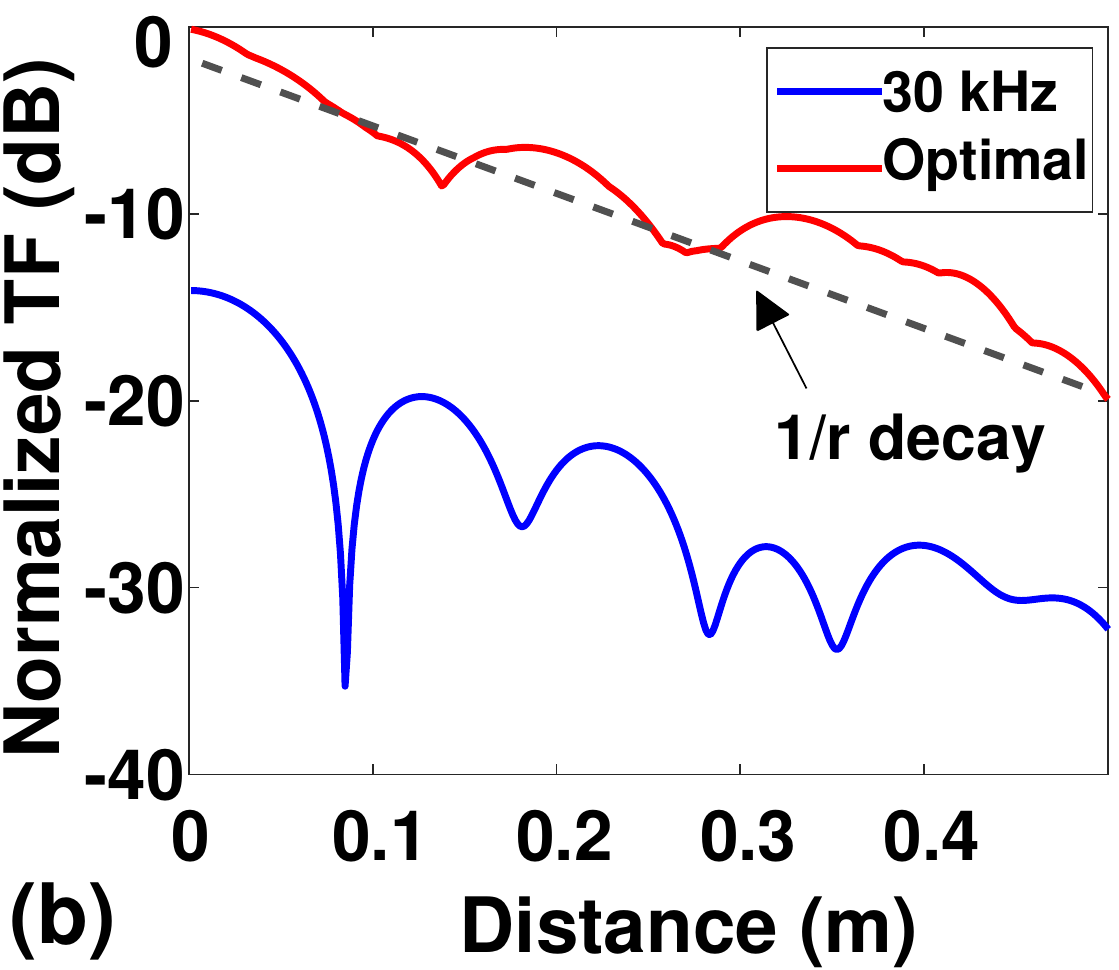}
    \caption{Measured power transmission in a 2~mm-thick stainless steel plate using Lamb waves: (a) power attenuation over long distances (0-1.5 m) in the 20-70~kHz range; (b) attenuation versus distance $r$ for both a randomly-selected frequency (30~kHz) and the distance-dependent optimal frequency.}
    \label{fig:power_trans}
\end{figure}

Guided ultrasound waves are also suitable for long-range wireless power and data transfer within structures~\cite{shaik2020self}. The resulting ultrasound channels are highly frequency-selective due to multi-path propagation, which generates patterns of constructive and destructive interference (known as ``slow fading''). For example, the measurements in Fig.~\ref{fig:power_trans}(a) show complex frequency- and distance-dependent power transmission patterns even in a simple structure (a uniform metal plate). Fig.~\ref{fig:power_trans}(b) shows that operating at the distance-dependent optimum frequency $f_{opt}(r)$ can significantly increase the available power (by $>15$~dB in this case). Also, the available power decays slowly with distance ($\propto 1/r$), as expected for guided waves in 2-D, thus enabling efficient long-distance power and (low-speed) data transfer. However, $f_{opt}(r)$ can change with time due to both environmental and structural changes (bending, temperature fluctuations, etc.). In earlier work, we addressed this challenge by developing a near-maximum power point tracking (nMPPT) algorithm that allows each node to track its own optimal transmission frequency~\cite{shaik2020self}.

\subsection{Ultrasonically-Coupled Network Architecture}
\begin{figure*}[htbp]
	\centering
	\includegraphics[width = 0.8\textwidth]{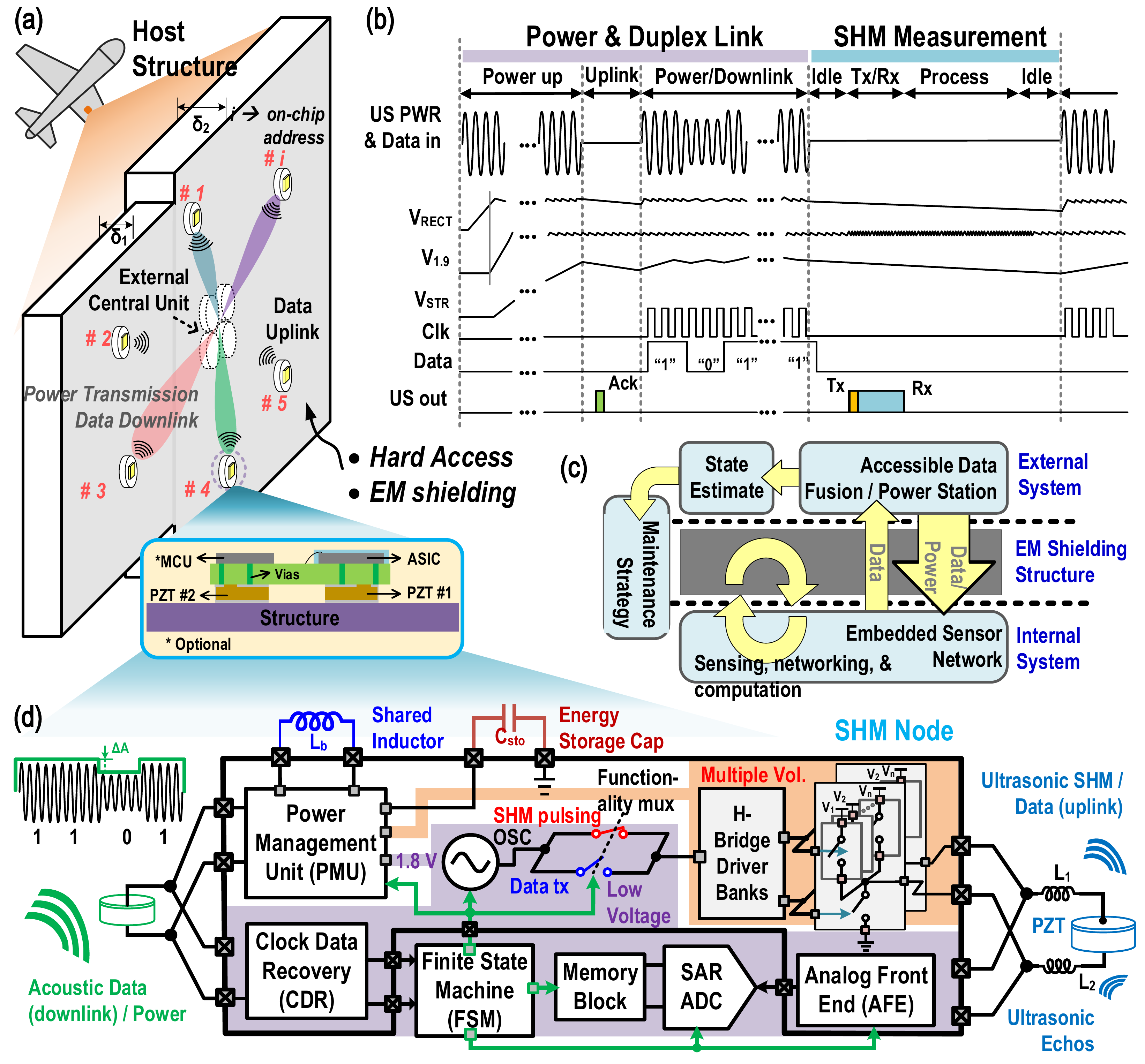}
	\caption{(a) Overview of the proposed ultrasonically-coupled wireless network for structural health monitoring (SHM). (b) High-level timing sequence of the combined ultrasonic power/data transfer and measurement cycle. (c) Block diagram of a closed-loop maintenance strategy enabled by a wireless SHM network deployed within a hard-to-access internal region. (d) Block diagram of the proposed miniature wireless SHM node. All the blocks shown inside the solid black line are integrated within the custom IC, while the others are implemented within an off-the-shelf MCU.}
	\label{fig:isc_overview_a}
\end{figure*}

The overall design of a self-optimizing ultrasonically-coupled network for an emerging SHM application, namely monitoring hard-to-access areas of a structure, is shown in Fig.~\ref{fig:isc_overview_a}(a). A wired central unit (known as the hub) provides access to the external world. The hub delivers ultrasonic power to a distributed set of sensor nodes, and also maintains bidirectional data links with them. Each node contains a custom ASIC and microcontroller (MCU) for making local SHM measurements and transferring the results back to the hub for further processing. Fig.~\ref{fig:isc_overview_a}(b) shows important waveforms during a typical measurement cycle. During the first part of the cycle, nMPPT is used to find $f_{opt}(r)$ and enough power is delivered to power up the node. Next, the node sends an acknowledge (ACK) signal via the data uplink (node to hub). On receiving ACK, the hub sends instructions to the node via the data downlink (hub to node), for which bits are modulated on the power carrier. The node then uses these instructions to set its parameters (e.g., operating frequency, pulse length, etc.) and carries out either a transmission- or reflection-type SHM measurement. Finally, it uses the uplink to transmit acquired data back to the hub and returns to an idle (or sleep) state. The hub integrates data from multiple nodes to develop estimates of the current state of the structure (e.g., maps of stress distribution or damage locations). Such estimates can then be used by human operators and/or AI algorithms to drive maintenance decisions, as shown in Fig.~\ref{fig:isc_overview_a}(c).

\section{Chip Design}
\label{sec:chip_design}
A highly-integrated ASIC is key for miniaturizing the proposed ultrasonically-coupled SHM sensor nodes (thus enabling deployment on non-planar surfaces) and also reducing their power consumption (thus enabling sparse sensor networks coupled via long-range links). Fig.~\ref{fig:isc_overview_a}(d) shows a block diagram of the proposed custom SHM IC, which is integrated together with an off-the-shelf ultra-low-power MCU to realize autonomous ultrasonically-coupled SHM sensor nodes. The chip is interfaced to two PWAS: the first (shown on the left) is used for acoustic power and data downlink from the central hub, while the second (shown on the right) is used for SHM measurements and also data uplink to the hub. Both links are designed to operate at programmable frequencies within the 100-500~kHz range to ensure single- or dual-mode Lamb wave propagation within a variety of structures (e.g., metal or carbon-fiber composite plates of different thicknesses).

The main components of the chip include the power management unit (PMU), clock and data recovery (CDR), SHM transmitter, load-shift keying (LSK)-based data transmitter, and the analog front-end (AFE) of the SHM receiver. The MCU contains the clock generator (which determines the operating frequency), a finite state machine (FSM) for sequencing the measurement, an ADC for digitizing the output of the SHM receiver, and on-chip memory (SRAM) for storing the results. In the next few sub-sections, we describe the major building blocks of the proposed IC in more detail.

\subsection{Active Rectifier}
\begin{figure}[htbp]
	\centering
	\includegraphics[width=1\columnwidth]{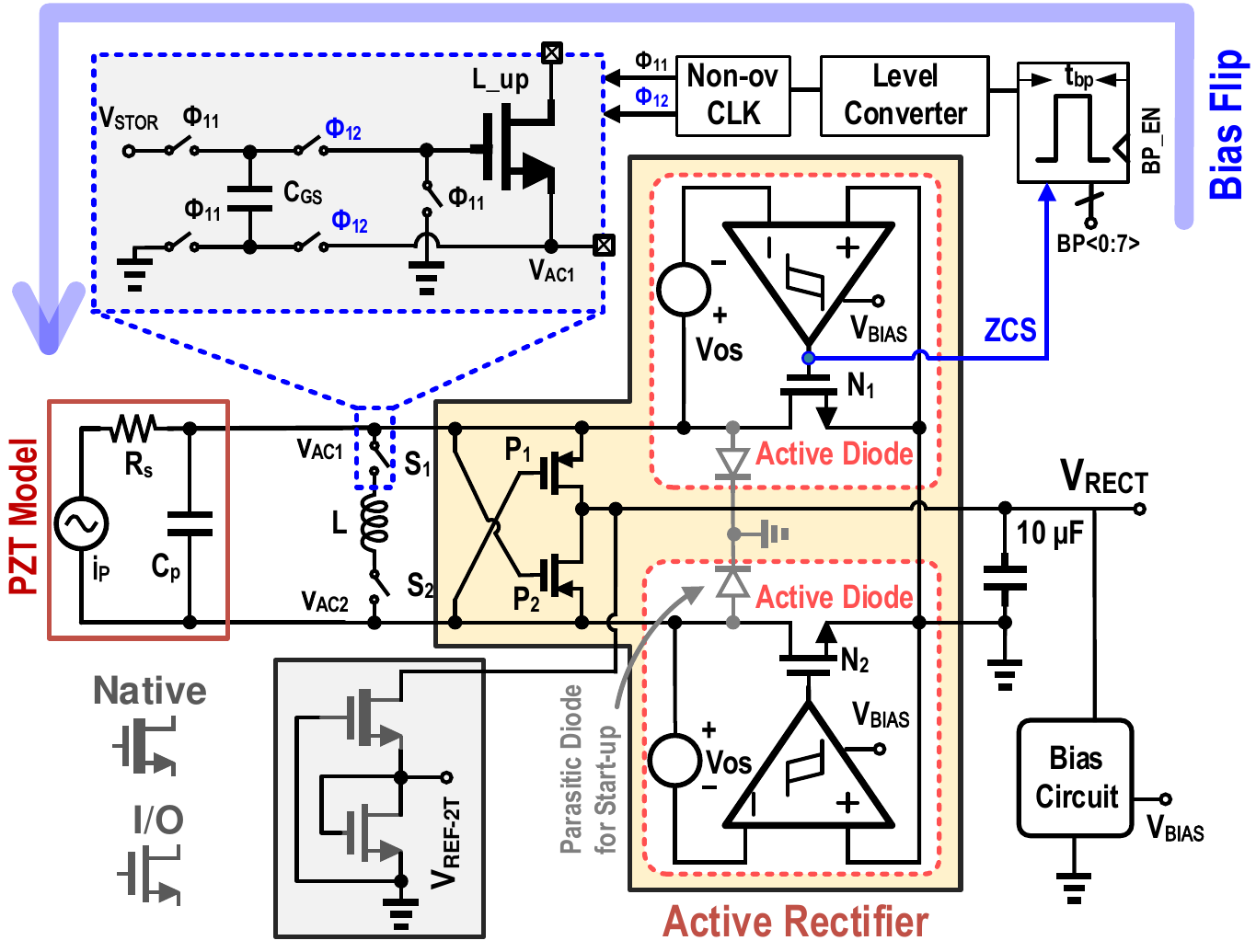}
	\caption{Block diagram of the AC-DC converter (active rectifier).}
	\label{fig:isc_ac_dc}
\end{figure}

Power management is a major function of the proposed ultrasonically-coupled SHM IC. The first component of the PMU is an AC-DC converter (rectifier) that converts the AC voltage across transducer (i.e., PWAS) \#1 into a DC voltage for recharging an energy storage capacitor. Fig.~\ref{fig:isc_ac_dc} shows the block diagram of the proposed rectifier. The core of the design is an active rectifier consisting of two PMOS switches (P$_1$ and P$_2$) and two NMOS switches (N$_1$ and N$_2$). The former are directly driven by the AC voltage across the transducer ($v_{AC1}-v_{AC2}$), while the latter are driven by hysteretic push-pull comparators to minimize i) voltage drop during the ON period; and ii) reverse current flow during the OFF period. Each comparator internally generates an adaptive offset voltage $V_{OS}$ to compensate for its own turn-OFF time delay, thus minimizing reverse current flow~\cite{erfani20191}. Note that only OFF-time delay compensation is implemented since reverse current directly degrades voltage and power conversion efficiency (VCE and PCE, respectively), while the ON-time delay only affects the conduction time of the switches.

\begin{figure}[htbp]
	\centering
	\includegraphics[width=1\columnwidth]{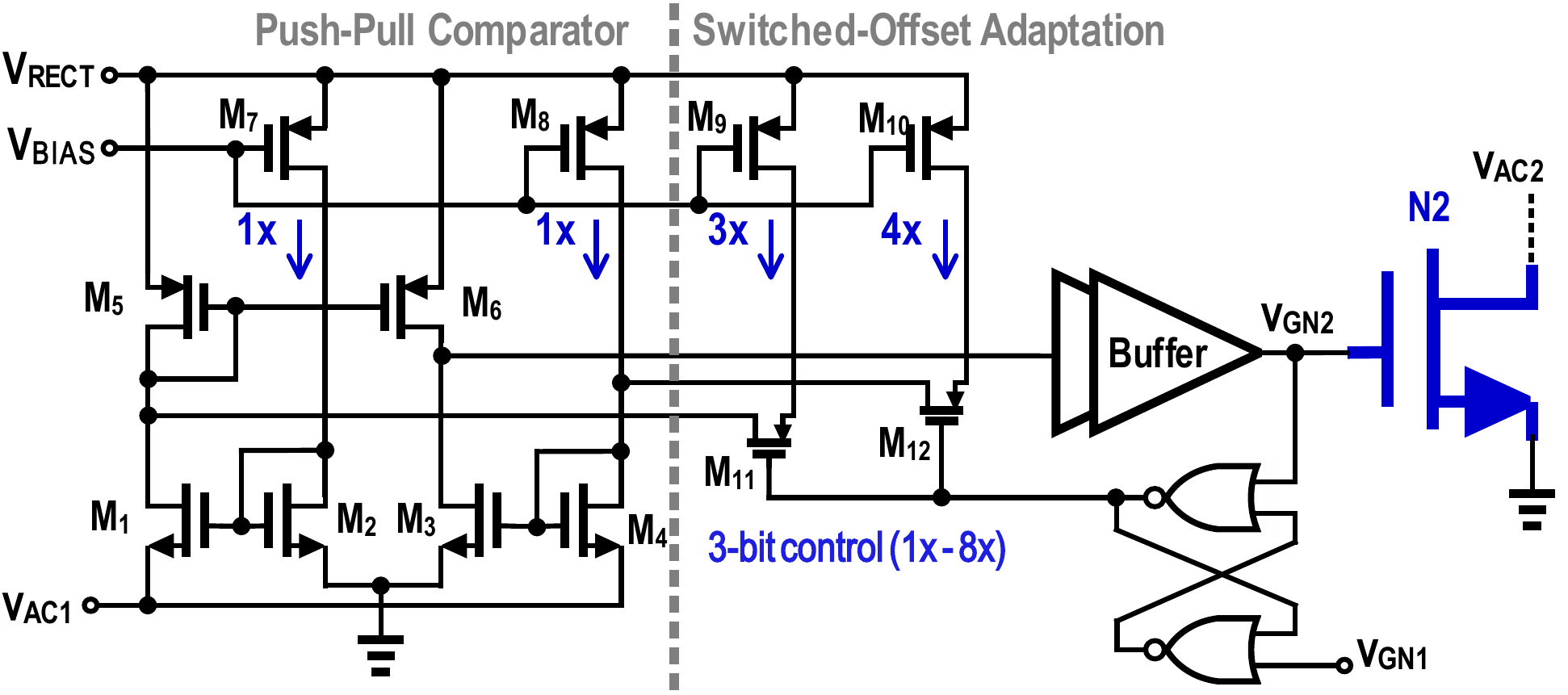}
	\caption{Schematic of the push-pull hysteretic comparator with switched-offset delay compensation used within the rectifier.}
	\label{fig:isc_comp}
\end{figure}

Fig.~\ref{fig:isc_comp} shows one of the comparators in more detail. The design compares $v_{AC1}$ (or $v_{AC2}$) with ground to generate the gate control signal $v_{GN2}$ (or $v_{GN1}$). The input current mirrors M$_1$-M$_4$ are biased at $I_{BIAS}\approx 200$~nA (denoted $1\times$ in the figure) by $V_{BIAS}$, which is generated by a PTAT current reference as shown in Fig.~\ref{fig:isc_ac_dc}. When $v_{GN2}$ goes low, it resets the SR latch, which turns on two offset currents (with nominal values of $3\times$ and $4\times$) that generate the required offset $V_{OS}$. The latch provides de-glitching and also ensures that the offset currents turn off when the complementary gate signal $v_{GN1}$ goes high. The optimal value of $V_{OS}$ is load-dependent, and can be optimized via 3-bit control of the offset currents.

Since the comparators are powered by the rectified output $V_{RECT}$, they are not available during ``cold-start'' conditions when the output capacitor is completely discharged. In this case, the parasitic drain-substrate diodes of $N_1$ and $N_2$ provide rectification (with lower efficiency), as shown in Fig.~\ref{fig:isc_ac_dc}.

In a conventional full-bridge rectifier, each diode contributes a forward voltage drop $V_{D}\approx 0.7$~V, such that the output DC voltage $V_{RECT}\leq V_{P}-2V_{D}$ where $V_P$ is the amplitude of the AC voltage across the transducer.  Moreover, the transducer has to charge its internal capacitance $C_P$ on every cycle, which wastes power. Fundamentally, this is because $v_{P}$ is out of phase with the transducer current $i_{P}$ due to the capacitive nature of the transducer impedance. The proposed rectifier uses ``bias-flip'' switches\footnote{This approach is also known as parallel-SSHI, where SSHI stands for ``synchronized switch harvesting on inductor''.} to improve the output power available from the transducer~\cite{ramadass2009efficient}. Specifically, switches $S_1$ and $S_2$ are placed in series with an off-chip shunt inductor $L$ and turned on when $i_{P}$ crosses zero. The inductive voltage $v=L(di/dt)$ then quickly flips the polarity of $v_P$, which reduces the energy loss caused by charging/discharging $C_P$. The resulting transducer voltage and current ($v_{P}$ and $i_{P}$) are nearly in phase (resembling a resistive impedance), which maximizes the available power. A bootstrap driver (shown in the inset of Fig.~\ref{fig:isc_ac_dc}) is used to create a floating gate-source voltage for controlling S$_1$ and S$_2$ through the AC cycle. During phase $\phi_{11}$, the DC voltage on the storage capacitor ($V_{STOR}$) is stored on $C_{GS}$. During phase $\phi_{12}$, the stored voltage is added to $V_{AC}$ to set the gate voltage. The necessary switches use dynamic body biasing to ensure that their parasitic diodes remain OFF. Switch timing for the proposed bias-flip circuit is controlled by a feedback loop that digitally adjusts the pulse width $t_{bp}$ (8-bit control) to ensure zero-current switching (ZCS).

\begin{figure}[htbp]
	\centering
	\includegraphics[width=0.8\columnwidth]{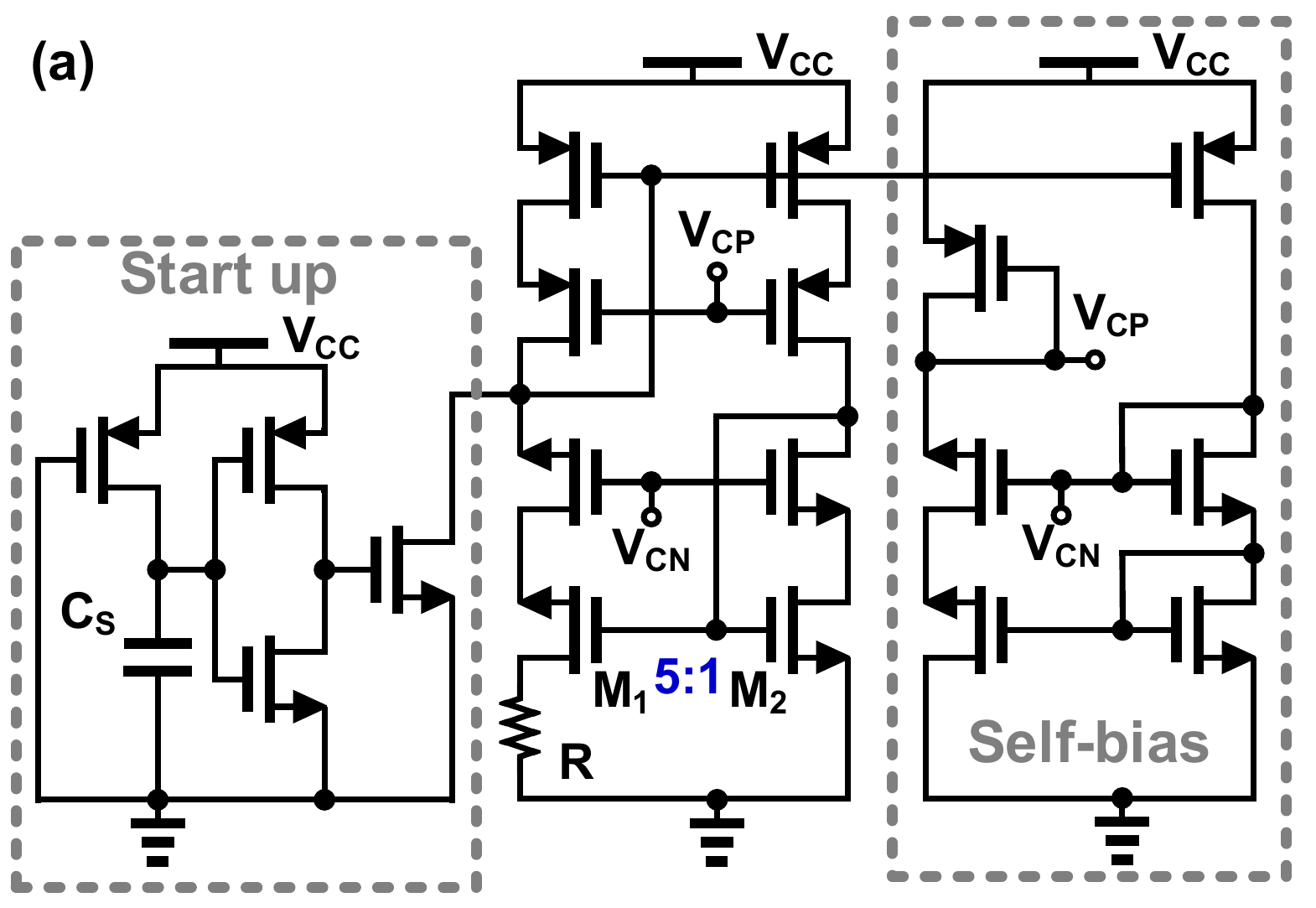}
	\includegraphics[width=0.7\columnwidth]{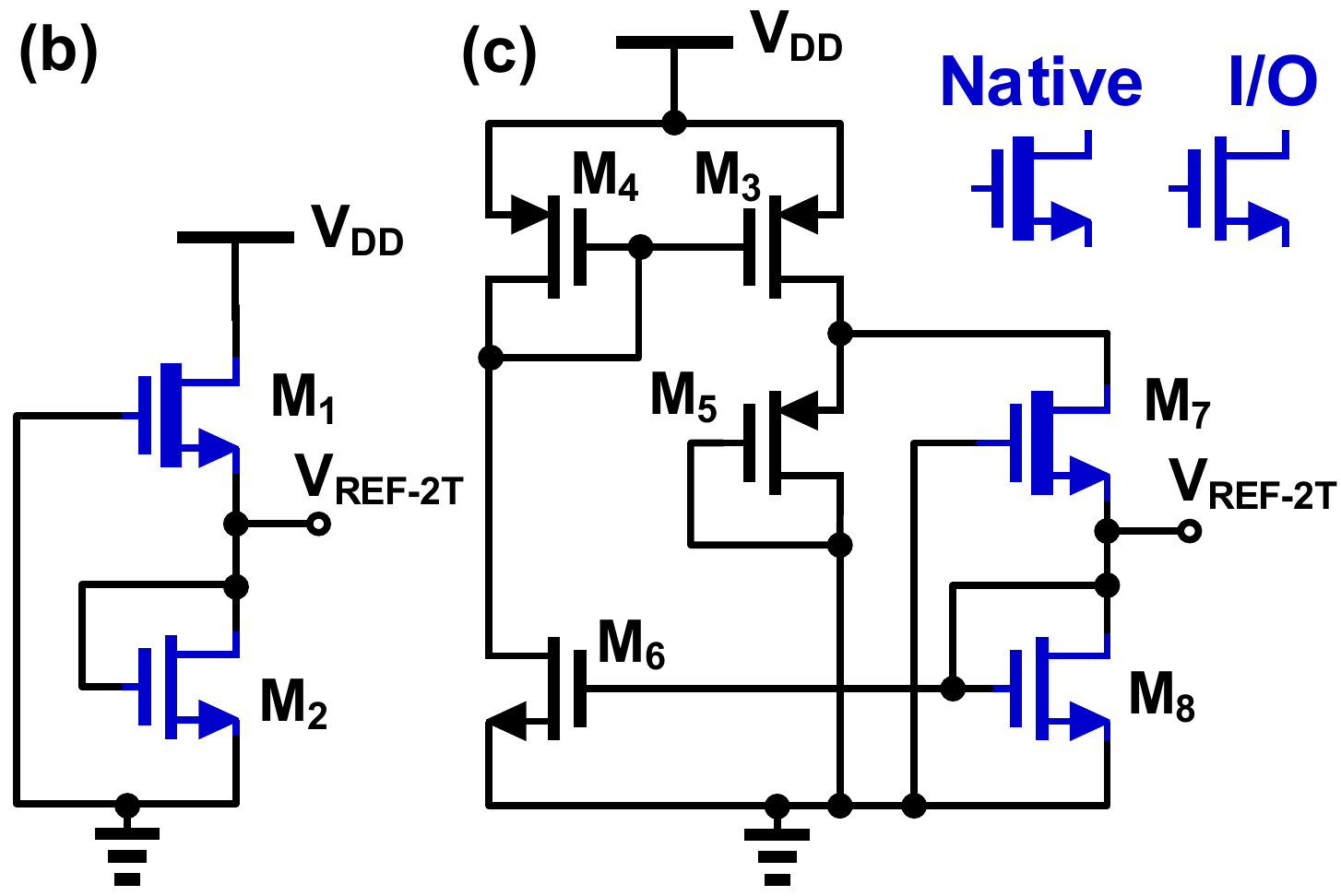}
	\caption{Schematics of the on-chip current and voltage reference circuits: (a) constant-$G_m$ current reference with start-up circuit; (b) basic 2T voltage reference; (c) modified 2T voltage reference with improved supply regulation.}
	\label{fig:isc_ref}
\end{figure}

The unregulated output DC voltage $V_{RECT}$ is used to power two bias generator circuits. The first is a fully-cascoded constant-$G_{m}$ reference with a nominal output current of $I_{BIAS}=200$~nA, shown in Fig.~\ref{fig:isc_ref}(a). The second uses a two-transistor (2T) reference~\cite{seok2012portable} to generate a PVT-robust voltage. A basic 2T reference using native and I/O transistors is shown in Fig.~\ref{fig:isc_ref}(b). We modified this design to improve power supply regulation, as shown in Fig.~\ref{fig:isc_ref}(c). The reference current $I_{REF}$ through M$_8$ is mirrored to M$_6$ and then M$_3$ with a ratio $1:m$, $m>1$. The diode-connected device M$_5$ then carries a current $(m-1)I_{REF}$, ensuring that $V_{DS}$ of M$_7$ becomes nearly independent of $V_{DD}$. The output voltage $V_{REF\_2T}\approx 300$~mV, which is similar to the original design.

\subsection{DC-DC Converter}
\begin{figure}[htbp]
	\centering
	\includegraphics[width=0.85\columnwidth]{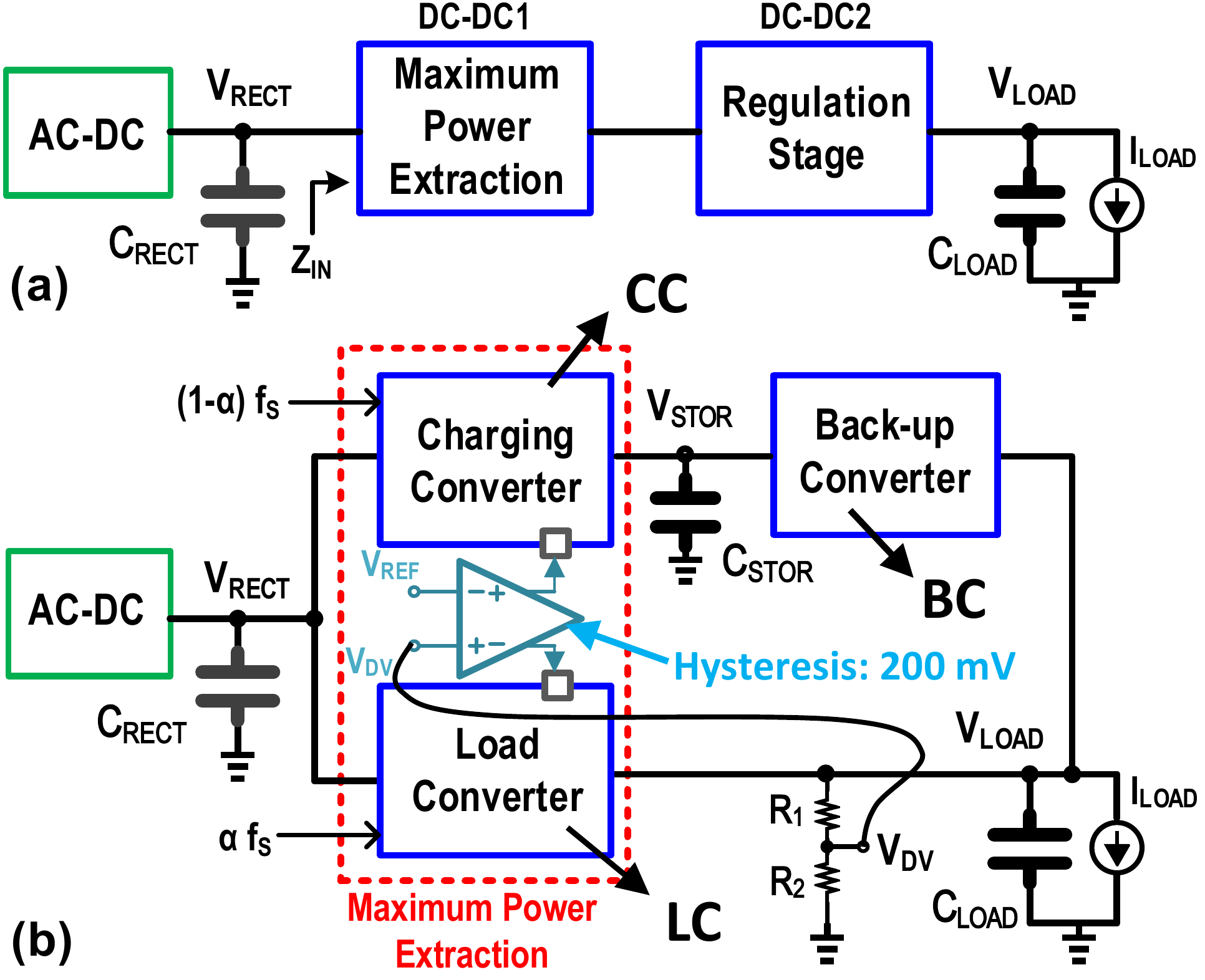}
	\caption{Architecture of the DC-DC converter: (a) conventional, and (b) proposed. The load is represented as a current source $I_{LOAD}$.}
	\label{fig:isc_dc_dc_architecture}
\end{figure}

The DC-DC converter transforms the unregulated rectifier output $V_{RECT}$ into two regulated outputs: $V_{STOR}$ (nominally 3.3~V) and $V_{LOAD}$ (nominally 2.0~V). The former is stored on a large energy reservoir $C_{STOR}$ (a super-capacitor in this case), while the latter powers the rest of the sensor node. Conventionally, these voltages are generated by two DC-DC converters in series, as shown in Fig.~\ref{fig:isc_dc_dc_architecture}(a). The first converter ensures maximum power point tracking (MPPT) by adjusting its input impedance to ensure maximum power transfer from the rectifier (and ultimately the ultrasound transducer), while the latter regulates the load voltage $V_{LOAD}$. Unfortunately, this series configuration suffers from reduced power efficiency, since the PCE of the cascaded converters is the product of the individual efficiencies, i.e., $\eta_{tot}=\eta_{1}\eta_{2}$.

By contrast, here we propose a dual-path architecture~\cite{bandyopadhyay2012platform} in which two boost converters - the charging converter (CC) and the load converter (LC) - are placed in parallel and adaptively selected by the current value of $V_{LOAD}$, as shown in Fig.~\ref{fig:isc_dc_dc_architecture}(b). The LC delivers power to the load, while the CC allows excess power from the rectifier to be stored in $C_{STOR}$ for later use. In addition, a back-up buck converter (BC) is used to charge the load from $V_{STOR}$ if the latter requires more power than is currently available from the rectifier. The system can transition between these states on every switching cycle based on the current value of $V_{LOAD}$. For this purpose, several resistively-divided versions of $V_{LOAD}$ (denoted by $V_{DV}$) are compared with a reference voltage $V_{REF}$ from the modified 2T circuit.

\begin{figure}
    \centering
    \includegraphics[width=1\columnwidth]{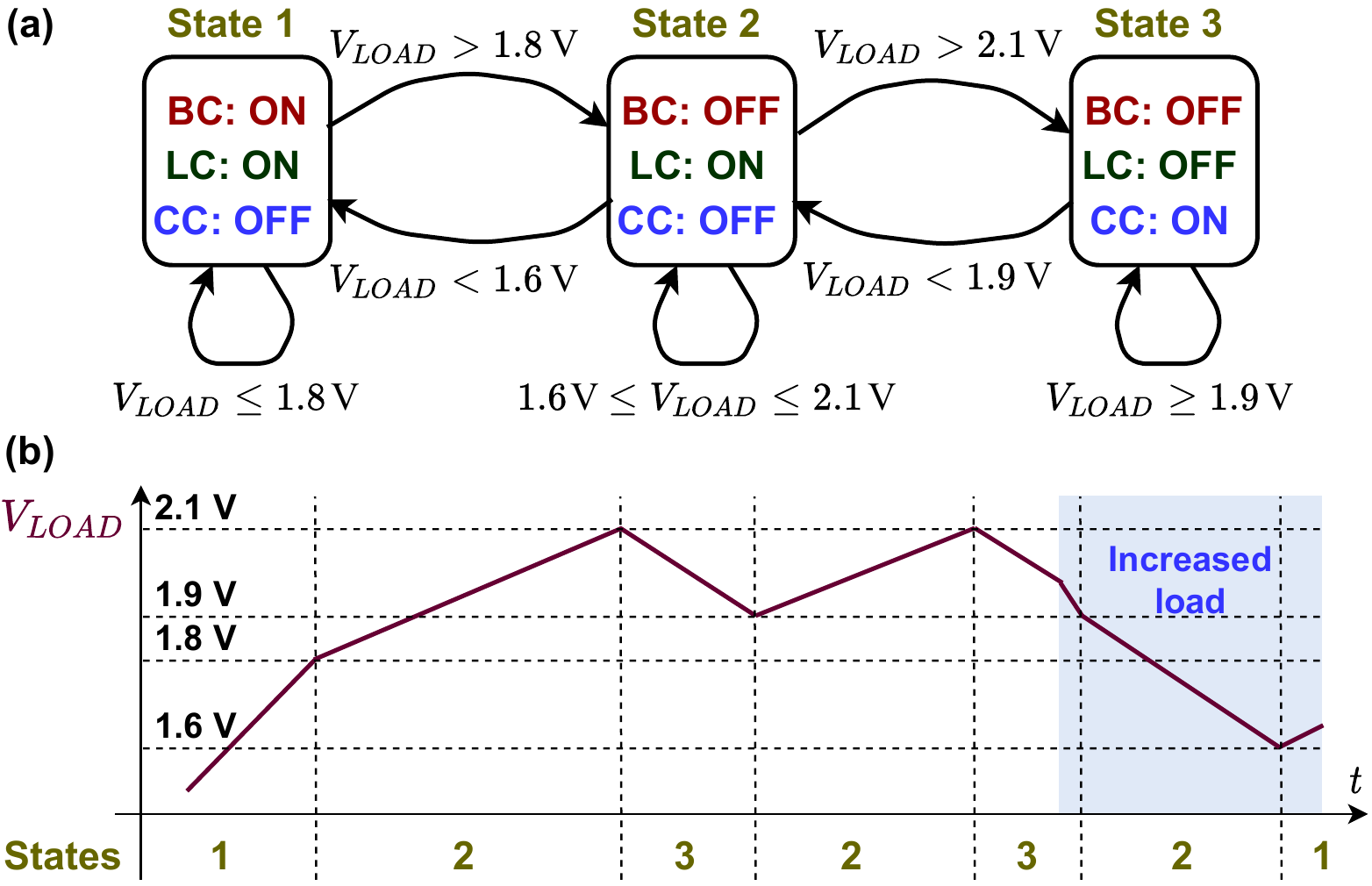}
    \caption{(a) State transition diagram of the proposed dual-path DC-DC converter; and (b) typical load voltage $V_{LOAD}$ for a typical load (unshaded region) and a heavy load (shaded region).}
    \label{fig:dc_dc_states}
\end{figure}

The resulting state transition diagram for the DC-DC converter can be summarized as shown in Fig.~\ref{fig:dc_dc_states}(a). Under normal conditions, the system cycles between states 2 and 3, resulting in a peak-to-peak ripple of $V_{H}=200$~mV around the nominal load voltage of 2.0~V as shown in Fig.~\ref{fig:dc_dc_states}(b); here $V_{H}$ is the amount of hysteresis between the switching thresholds for states 2 and 3 (2.1~V and 1.9~V, respectively). However, when the harvested power is insufficient for the load and $V_{LOAD}$ drops below 1.6~V, the system transitions to state 1 (see Fig.~\ref{fig:dc_dc_states}(b)), where both the BC and LC are turned on to rapidly recharge $C_{LOAD}$. The switching thresholds for states 1 and 2 (1.8~V and 1.6~V, respectively) are also offset by $V_{H}=200$~mV. The resulting fluctuations in $V_{LOAD}$ are acceptable for this application, but can be reduced by using a higher-resolution divider to set a smaller value for $V_{H}$.  

Let us denote the probability that the system is in state 2 by $p(2)=\alpha$ ($0<\alpha<1$). Clearly, the probability that the system is in either of the other two states is $p(1)+p(3)=(1-\alpha)$. Also, let us denote the PCE of the LC and CC (assumed to be equal for simplicity) as $\eta_1$, and the PCE of the BC as $\eta_2$. The PCE during state 2 is simply $\eta_1$ since only the LC is operating. On the other hand, states 1 and 3 together transfer energy to the load in two steps ($V_{RECT}\rightarrow V_{STOR}\rightarrow V_{LOAD}$), resulting in a effective PCE of $\eta_1\eta_2$. Thus, the average end-to-end PCE of the proposed DC-DC converter is given by
\begin{equation}
\eta_{tot}=\eta_1\times\alpha+\eta_1\eta_2\times(1-\alpha).
\label{eq:eta_tot}
\end{equation}

The availability of an upload data link (node to hub) allows the hub to regulate its output power level such that the harvested power is approximately equal to that consumed by the load. In this case the CC and BC are mostly inactive, i.e., $\alpha\approx 1$. As a result, (\ref{eq:eta_tot}) simplifies to $\eta_{tot}\approx \eta_1$, which is likely to be significantly higher than the cascaded converter architecture. More generally, the system may receive more harvest more power than needed by the load, in which case it cycles between states 2 and 3 as described earlier. In this case, $p(1)\approx 0$, such that $p(3)=(1-\alpha)$ and the effective duty cycles of the LC and CC are $\alpha$ and $(1-\alpha)$, respectively.

\begin{figure}[htbp]
	\centering
	\includegraphics[width = 1\columnwidth]{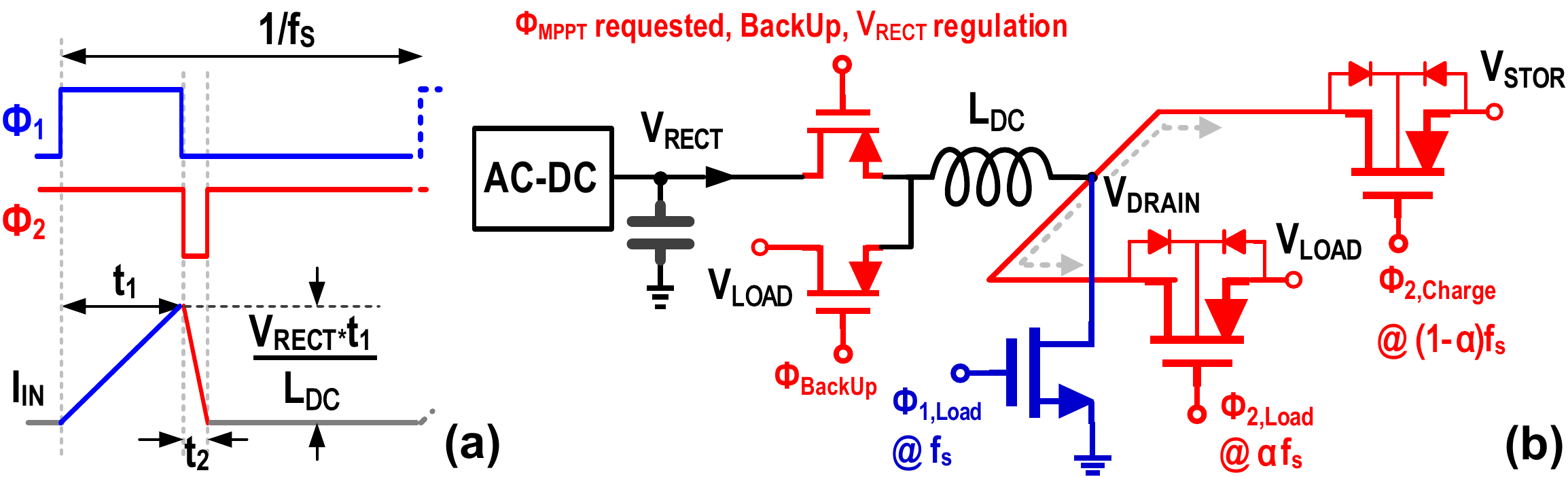}
	\caption{(a) Current waveform in DCM mode for the DC-DC converter. (b) Inductor time-sharing scheme used by the converter.}
	\label{fig:isc_dc_dc_zin}
\end{figure}

All three converters operate in discontinuous conduction mode (DCM), which allows them to use a single time-shared off-chip inductor $L_{DC}$. The input impedance $R_{IN}$ (as seen by the rectifier) for a single boost converter can be found by estimating the average inductor current $\overline{I_{IN}}$ per cycle. Given the pulse widths $t_1$ and $t_2$ for the two switches (as controlled by the clock phases $\Phi_1$ and $\Phi_2$, see Fig.~\ref{fig:isc_dc_dc_zin}(a)), the result is
\begin{align}
\overline{I_{IN}}& = \frac{1}{2}\left(t_1+t_2\right)\frac{V_{RECT}\cdot t_1}{L_{DC}}f_s~~\mathrm{and}\\
R_{IN}\equiv \frac{V_{RECT}}{\overline{I_{IN}}}& = \frac{2L_{DC}}{t_1^2f_s}\left(1+\frac{t_2}{t_1}\right)^{-1}\approx \frac{2L_{DC}}{t_1^2f_s},
\end{align}
where $f_s$ is the switching frequency (fixed at 50~kHz in our design) and the approximation is valid when $t_{2}\ll t_{1}$. Thus, $t_1$ can be controlled to adjust $R_{IN}$ and thus ensure MPPT.

The switch configuration used for inductor time-sharing is shown in Fig.~\ref{fig:isc_dc_dc_zin}(b). There are a total of five switches: two series switches for the CC and LC, respectively; one shunt switch shared by them; one series switch for the BC; and one series switch for implementing both MPPT and rectifier under-voltage protection. Under normal circumstances when $p(1)\approx 0$, either the LC or CC is always on (with duty cycles of $\alpha$ and $(1-\alpha)$, respectively). As a result, the average value of $R_{IN}$ is unchanged, which allows MPPT to be maintained:
\begin{equation}
R_{IN,av} \approx \left[\alpha R_{IN,LC}^{-1}+(1-\alpha)R_{IN,CC}^{-1}\right]^{-1}= \frac{2L_{DC}}{t_1^2f_s}.
\end{equation}

\begin{figure}[htbp]
	\centering
	\includegraphics[width = 1\columnwidth]{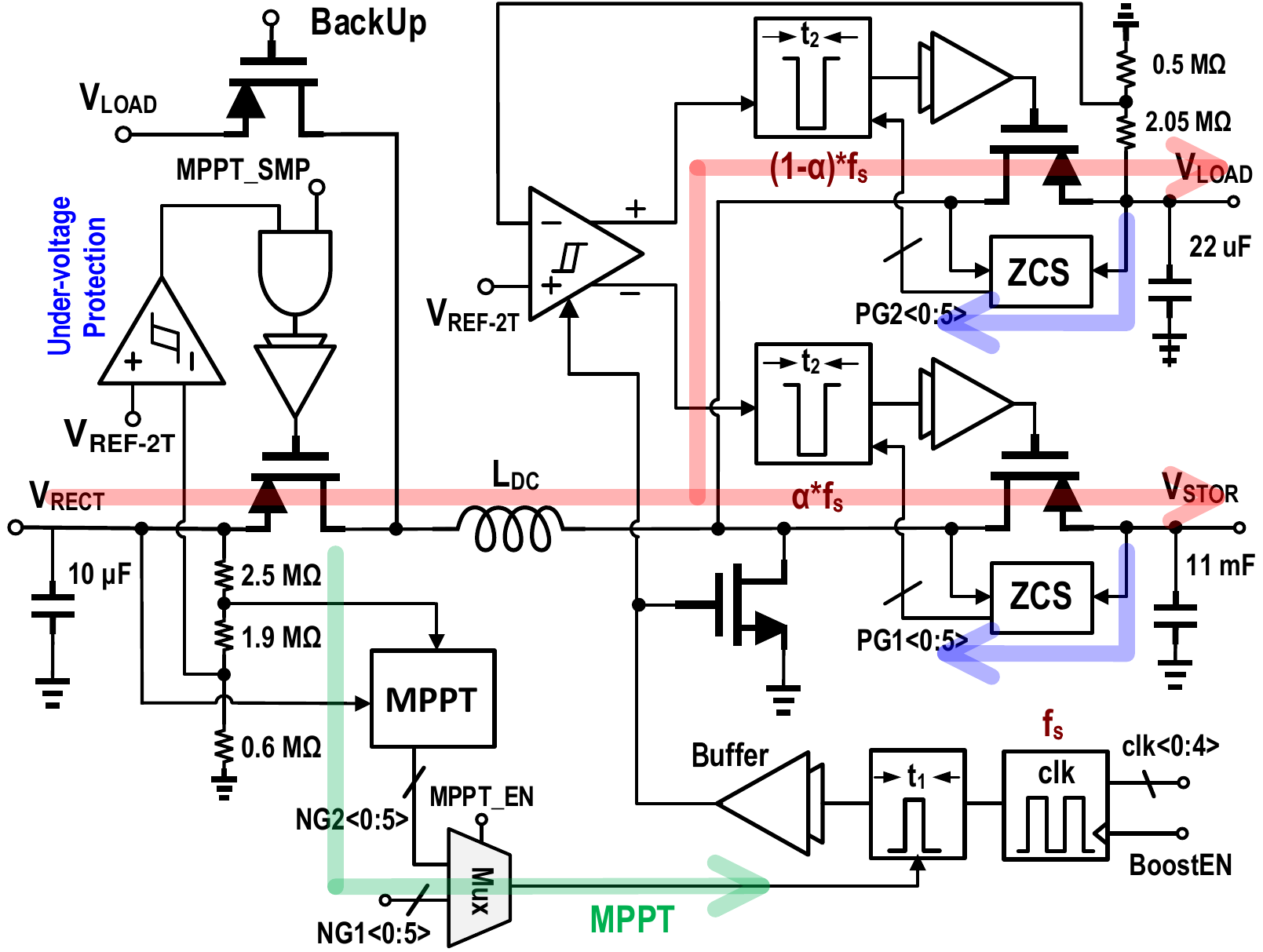}
	\caption{Block diagram of the DC-DC converter circuit, including the power flow path (orange), MPPT control loop (green), and ZCS control loops (blue).}
	\label{fig:isc_dc_dc}
\end{figure}

Fig.~\ref{fig:isc_dc_dc} shows a simplified view of how the DC-DC architecture was implemented on-chip. Two feedback loops are used to adapt the switch timings, as shown by the green and blue arrows in Fig.~\ref{fig:isc_dc_dc}. The green MPPT loop turns off the first series switch when the signal MPPT\_SMP goes high, thus disconnecting the DC-DC converter from the rectifier. A sample-and-hold (S/H) within the ``MPPT'' block then measures the open-circuit voltage $V_{RECT,0}$ of the rectifier. Note that a hysteretic comparator also turns off this switch when the loaded rectifier voltage $V_{RECT}$ drops below a pre-set threshold, thus providing under-voltage protection. During normal operation, the loop ensures MPPT by adjusting the duration of $\Phi_{1}$ (denoted by $t_1$), and thus $R_{IN}$, such that the loaded value of $V_{RECT}=0.5\times V_{RECT,0}$; this is generally a good approximation to the maximum power point. Adjustment is performed digitally using an 6-bit accumulator, and can be disabled by a control signal MPPT$\_$EN as shown in Fig.~\ref{fig:isc_dc_dc}. 

\begin{figure}[htbp]
	\centering
	\includegraphics[width = 0.8\columnwidth]{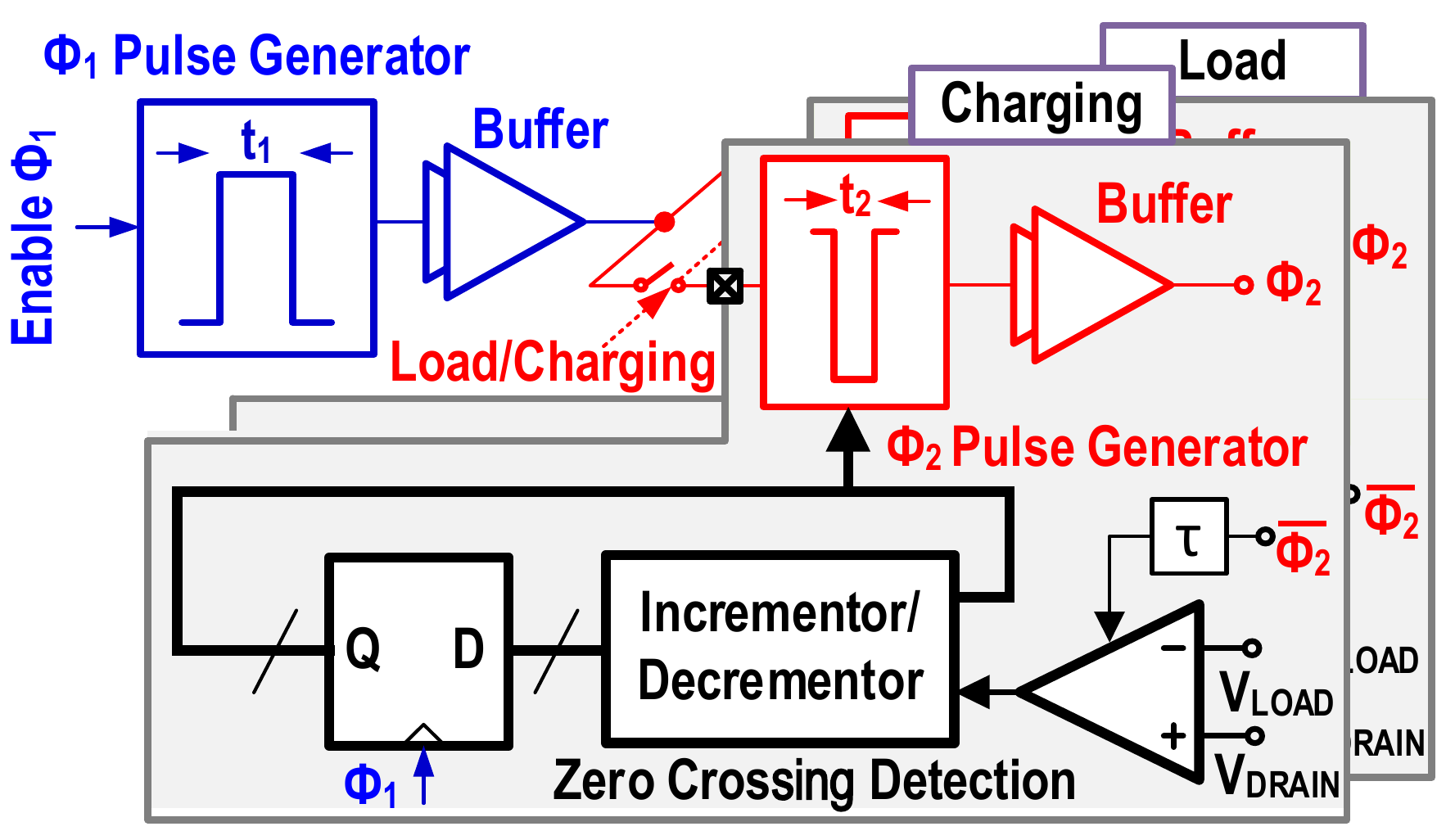}
	\caption{Schematic of the ZCS control loops for the boost converters.}
	\label{fig:zcs_loop}
\end{figure}

The second feedback mechanism consists of independent ZCS loops for the LC and CC. These blue loops are shown in more detail in Fig.~\ref{fig:zcs_loop}. They use similar 6-bit digital control circuits to adjust the individual durations of $\Phi_{2}$ (denoted by $t_2$) such that ZCS is obtained for the inductor current waveform, thus maximizing power efficiency. Finally, the present value of $V_{LOAD}$ is used to switch between the LC and CC as described earlier, thus ensuring load voltage regulation. 

\subsection{SHM Transmitter}
\begin{figure}[htbp]
	\centering
	\includegraphics[width=0.85\columnwidth]{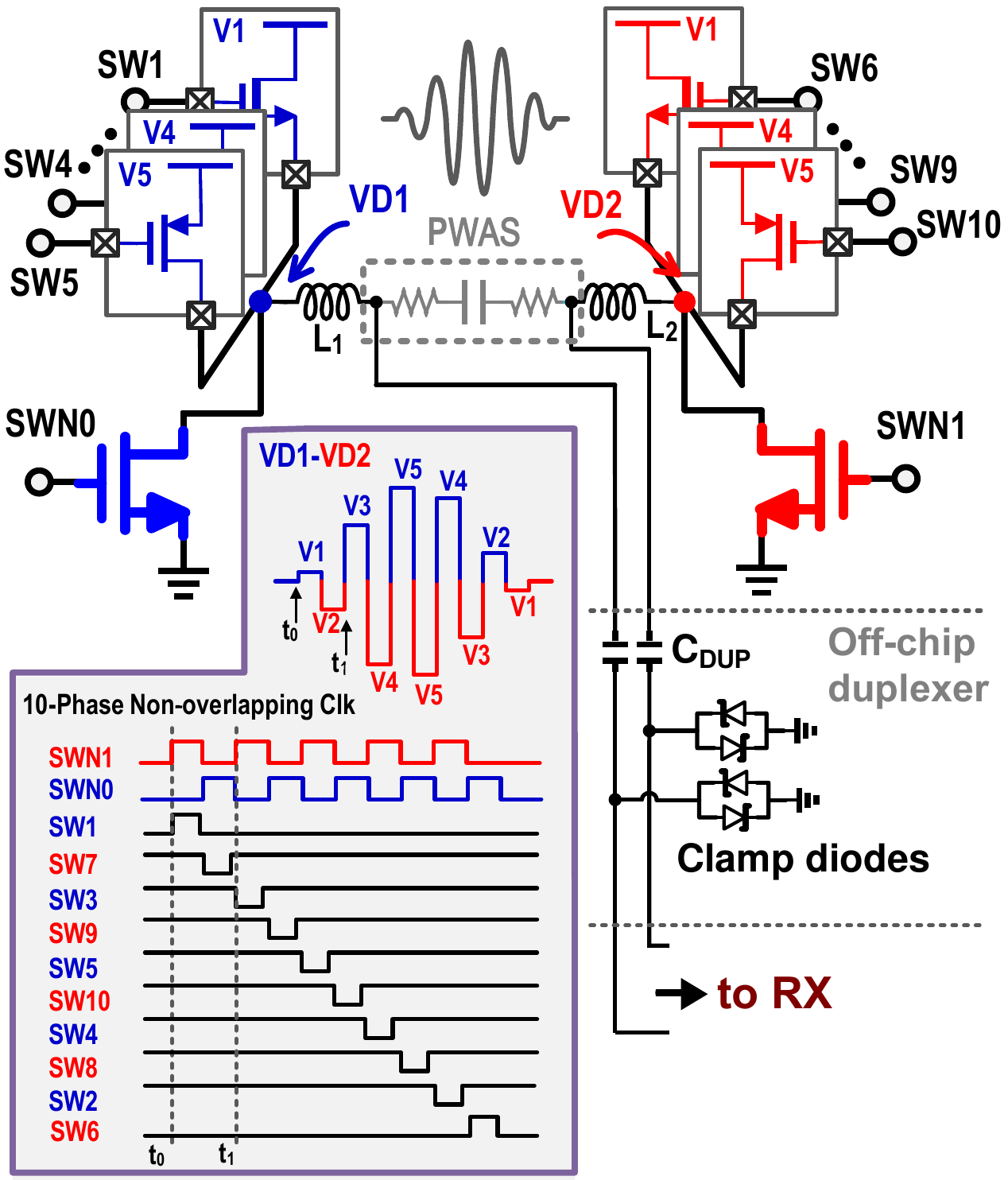}
	\caption{Block diagram of the SHM transmitter.}
	\label{fig:isc_tx_rx}
\end{figure}

Fig.~\ref{fig:isc_tx_rx} shows a block diagram of the transmitter (Tx) used for SHM measurements. The Tx generates pulses across PWAS \#2 with shapes that approximate a Hamming-windowed tone burst, as shown in the figure. As described in Section~\ref{sec:shm}, windowing ensures that the pulse is localized in the frequency domain (i.e., does not have significant side-lobes), which improves the accuracy of SHM measurements by avoiding the excitation of multiple propagating Lamb wave modes. The earlier (wired) SHM transceiver IC described in~\cite{tang2018programmable} used pulse-width modulation (PWM) to generate a close approximation to a Hamming-windowed pulse. However, PWM requires the use of a high-frequency clock (in the earlier design, $16\times$ higher than the operating frequency) to generate narrow pulses, which significantly increases overall power consumption of the IC. The new design eliminates this problem by dynamically switching between multiple power supply voltages during the pulse, thus directly controlling its amplitude on a cycle-by-cycle basis. The cycle period, and thus the center frequency $f_0$ of the pulse, is externally programmable via the clock frequency, resulting in a $-3$~dB excitation bandwidth of $1.30\times f_0/5\approx f_0/4$ for a five-cycle pulse, where the factor of $1.30$ arises from the Hamming window.

The transmitter circuit uses a H-bridge topology to ensure high power-efficiency and maximize the output signal swing (up to $\pm V_{DD}$), as shown in Fig.~\ref{fig:isc_tx_rx}. Small off-chip inductors $L_1$ and $L_2$ are placed in series with the load (i.e., the ultrasound transducer) to create a series $LRC$ circuit with quality factor $Q$, which suppresses harmonics and further boosts the steady-state voltage across the transducer by a factor of up to $Q$. The high-side PMOS switches of the H-bridge are split into five pairs ($SW1$--$SW10$). Each pair is supplied by a different DC supply voltage (denoted $V_1$--$V_5$), thus enabling the pulse envelope to be dynamically controlled using the switching sequence shown in the figure, which is generated on-chip from a single external trigger pulse. The necessary DC voltages are generated by on-chip switched-capacitor DC-DC converters, as described next.

\subsection{Switched-Capacitor DC-DC Converters}
A set of five parallel switched-capacitor converters use pulse-frequency modulation (PFM)~\cite{kwong200865} to efficiently generate the five regulated output voltages required by the transmitter, namely $V_1=0.36$~V, $V_2=0.89$~V, $V_3=1.9$~V, $V_4=2.87$~V, and $V_5=3.3$~V. These values were obtained by numerical optimization, with the target being the best possible five-level approximation to the desired SHM transmit pulse waveform (five cycles long, Hamming-windowed). Simulations show that the five-level pulse has 20~dB lower worst-case sidelobe level than a simple ``on-off'' tone burst, resulting in a peak-to-sidelobe ratio (PSL) of approximately $-32$~dB. 

\begin{figure}[htbp]
	\centering
	\includegraphics[width=1\columnwidth]{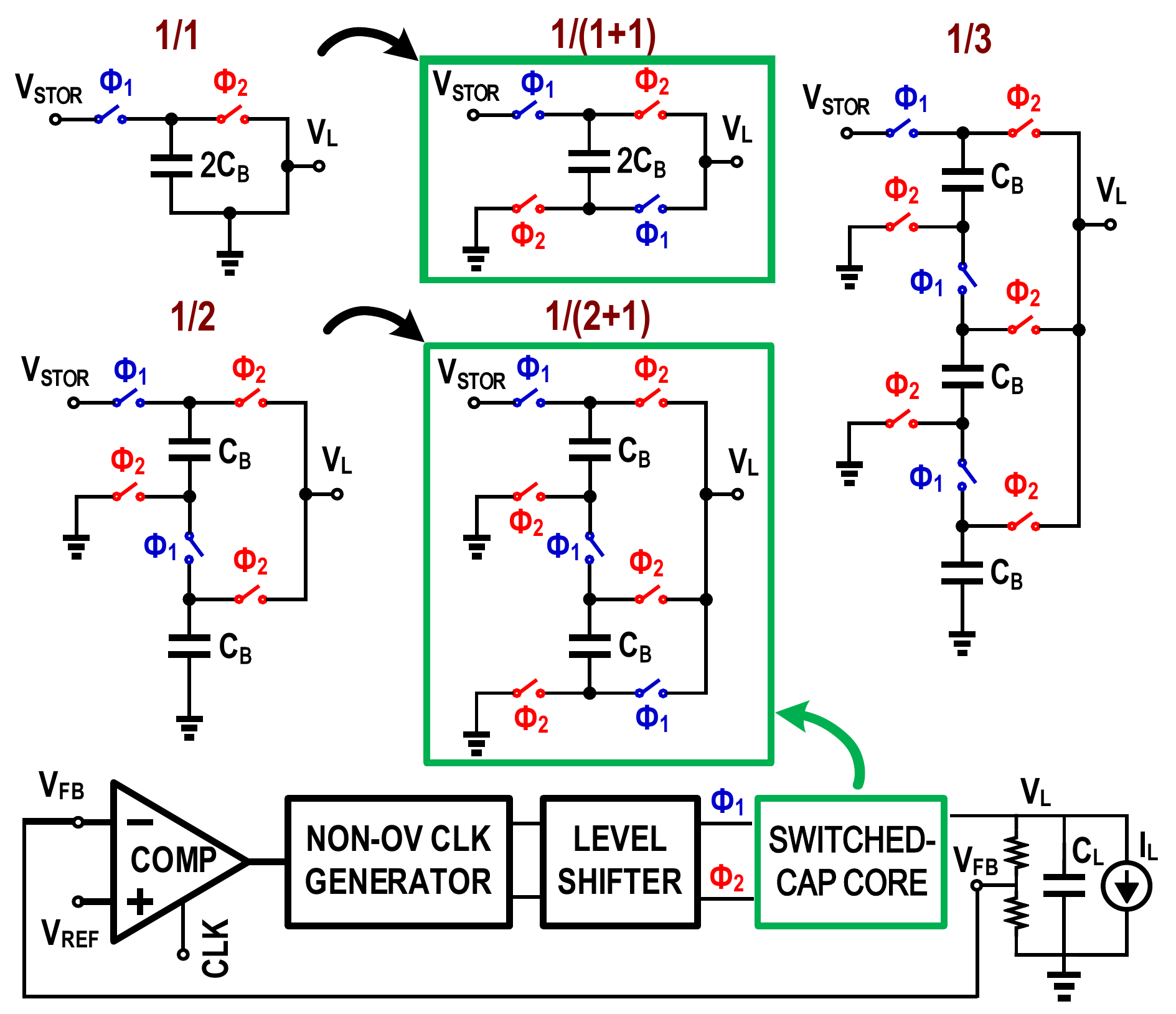}
	\caption{Block diagram of the switch configurations (top) and PFM feedback loop (bottom) used by the switched-capacitor DC-DC converters. Only the highlighted switch configurations were utilized in the final design. The 1/1 configuration (a single switch) is not shown for simplicity.}
	\label{fig:isc_switched_cap}
\end{figure}

Three of the five required DC levels were generated by 1/1 ratio converters: i) $V_{STOR}>3.3$~V $\rightarrow$ 3.3~V and 2.87~V; and ii) $V_{LOAD}>1.9$~V $\rightarrow$ 1.9~V. The two remaining voltage levels were obtained as follows: i) a 1/3 ratio converter to generate 0.89~V from $V_{STOR}$, and ii) a 1/2 ratio converter to generate 0.36~V from 0.89~V. The necessary switched-capacitor configurations are shown in Fig.~\ref{fig:isc_switched_cap}. The 1/1 converters consist of a single switch between the input ($V_{STOR}$ in the figure) and output ($V_{L}$), and are not shown for simplicity. For the 1/2 and 1/3 converters, we used the configurations highlighted in Fig.~\ref{fig:isc_switched_cap} (``1/(1+1)''and ``1/(2+1)'', respectively) rather than the alternative designs also shown in the figure, which require more switches and thus have lower efficiency. The ``1/(1+1)''and ``1/(2+1)'' configurations can be derived from basic 1/1 and 1/2 converter designs by adding a single pair of switches, as indicated by the arrows. 

A similar PFM-based voltage-regulation loop is used for each converter, as shown in Fig.~\ref{fig:isc_switched_cap}. The loop feeds back a divided version of the output voltage $V_L$ (denoted by $V_{FB}$) to a hysteretic comparator, which then enables/disables the local non-overlapping clock generator. A relatively low value of hysteresis (50~mV) was used to ensure low output voltage ripple, and thus accurately-shaped pulses. The on-chip capacitance used by the 1/2 and 1/3 converters was set to a relatively large value ($C_{B}=25$~pF) to ensure low output impedance. Finally, voltage droop during the pulse was minimized by using off-chip capacitors ($C_{L}=50$~nF) to store enough charge at each output node. 

\subsection{Transmit-Receive Switch}
Since the PWAS voltage can significantly exceed $V_{DD}$ (due to the voltage amplification provided by the resonant $LRC$ load), a simple off-chip transmit-receive switch (duplexer) was used to protect the input terminals of the SHM receiver. This circuit consists of small series capacitors $C_{DUP}$ and two sets of back-to-back diode clamps, as shown in Fig.~\ref{fig:isc_tx_rx}. In transmit mode, the diodes turn on (behaving as a short circuit), thus placing an effective capacitance $C_{DUP}/2$ in parallel with the PWAS and limiting the voltage across the receiver terminals to one diode drop (approximately $\pm 0.7$~V). The value of $C_{DUP}$ is chosen to be significantly smaller than the transducer capacitance $C_{P}$ to ensure that the added capacitance does not significantly degrade transmitter efficiency. In receive mode, the diodes turn off (behaving as an open circuit), thus allowing the received signals to pass into the receiver through $C_{DUP}$.

\subsection{Receiver}
\begin{figure}[htbp]
	\centering
	\includegraphics[width=1\columnwidth]{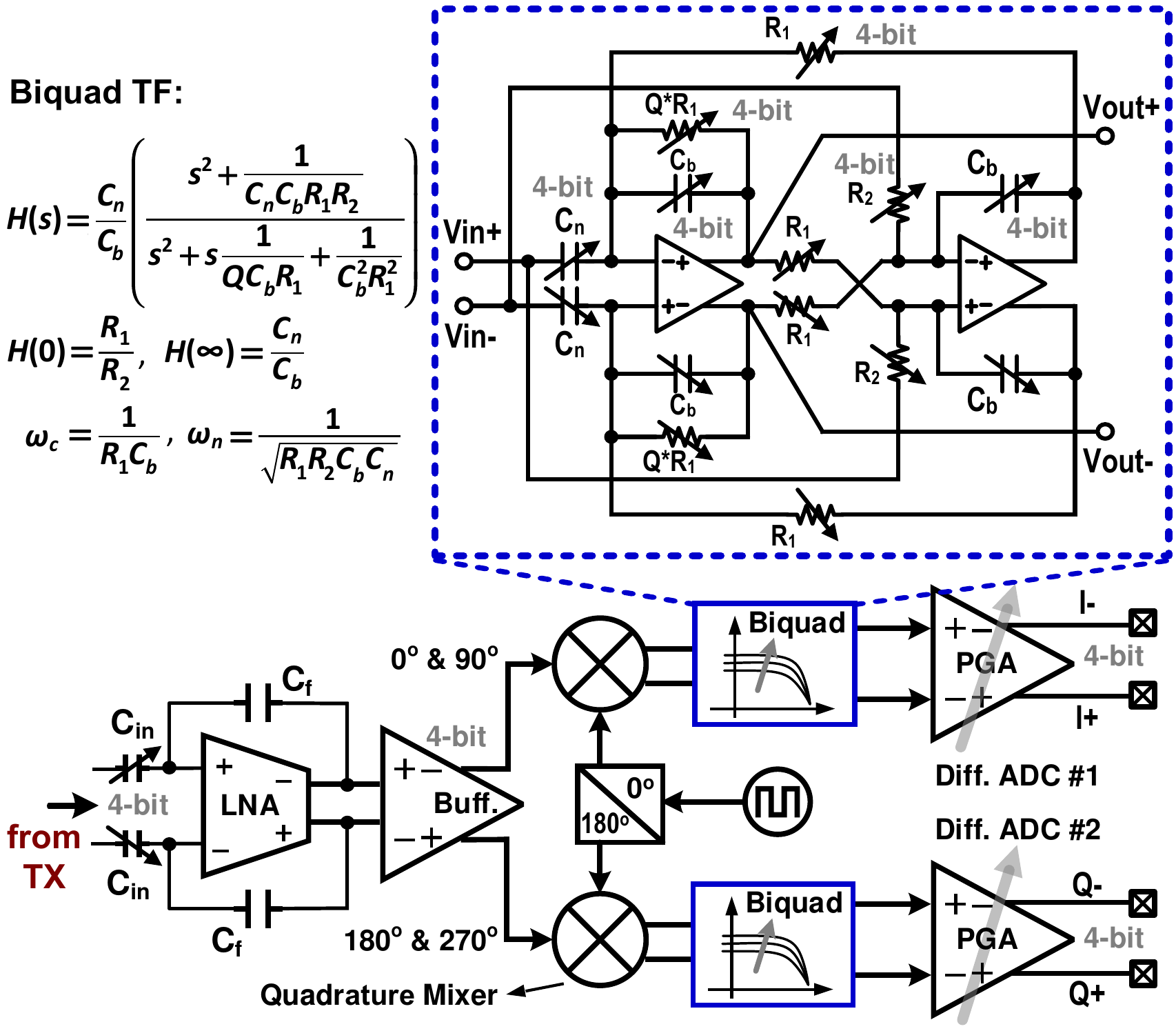}
	\caption{Block diagram of the SHM receiver. The fully-differential I/Q outputs are digitized off-chip by a two-channel differential ADC, as indicated.}
	\label{fig:isc_rx}
\end{figure}

The receiver design is shown in Fig.~\ref{fig:isc_rx}. The first stage is a fully-differential low-noise amplifier (LNA) based on a folded-cascode OTA topology. The gain of the LNA is set by capacitive feedback to $A_{v}=C_{in}/C_{f}$; here $C_{in}$ is digitally programmable via a 4-bit capacitor DAC. The presence of $C_{DUP}$ causes the gain to decrease to $A_{v}=C_{in,eff}/C_{f}$ where $C_{in,eff}=C_{in}C_{DUP}/\left(C_{in}+C_{DUP}\right)$, thus degrading the input-referred noise from its design value of $\sim$10~nV/Hz$^{1/2}$. To minimize such degradation, we set $C_{DUP} \gg C_{in,max}$. Thus, the condition $C_{P}\gg C_{DUP} \gg C_{in,max}$ is required for the proposed passive duplexer to work as intended. In our design $C_{P}\approx 100$~pF and $C_{in,max}=2$~pF, so we used $C_{DUP}=10$~pF to approximate $\sqrt{C_{P}C_{in,max}}$, which is a suitable value. Note that we have ignored the off-state capacitance $C_{d}$ of the diodes (which further attenuates the received signal) for simplicity; low-capacitance diodes ($C_{d}<1$~pF) were used to minimize additional signal attenuation.

\begin{figure*}[htbp]
	\centering
	\includegraphics[width = 0.86\textwidth]{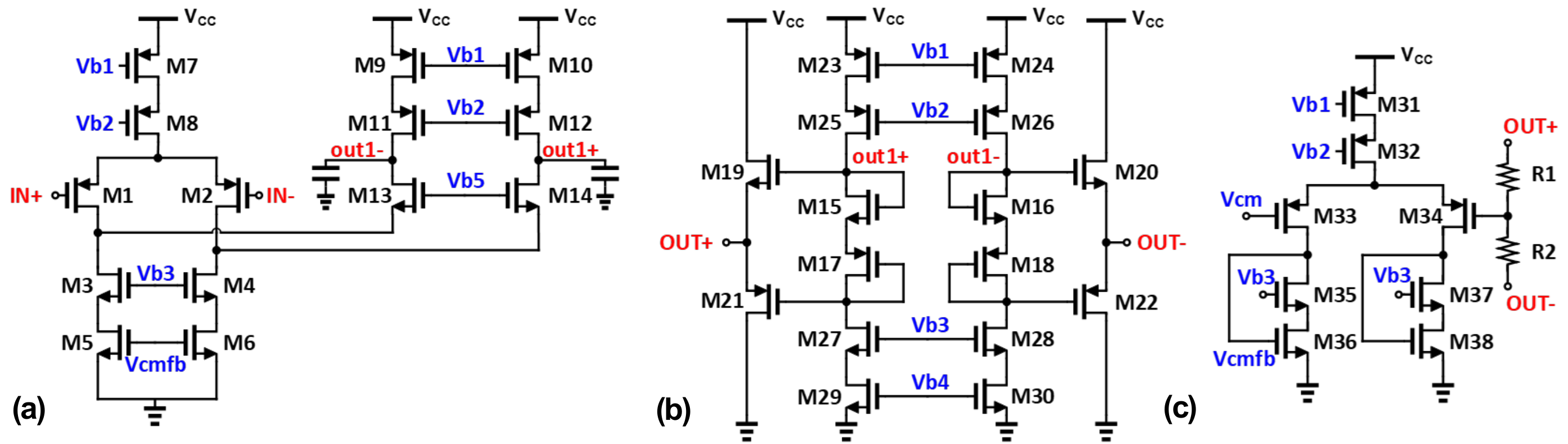}
	\caption{Schematic of the fully-differential op-amp used in the SHM receiver: (a) first stage, (b) second stage, (c) common-mode feedback (CMFB) circuit.}
	\label{fig:isc_opamp}
\end{figure*}

The rest of the receiver uses a standard super-heterodyne topology with fully-differential signal path to minimize even-order distortion and common-mode noise. Quadrature down-conversion (using passive double-balanced mixers) is used to down-convert received SHM signals to baseband, thus minimizing the sampling rate required by the MCU's ADC. The LNA output is buffered by a fully-differential folded-cascode op-amp (shown in Fig.~\ref{fig:isc_opamp}) before driving the mixers. The op-amp uses a nominal bias current of $I_{B}=2$~$\mu$A derived from a 1~$\mu$A constant-$G_m$ reference. The latter also generates the other op-amp bias voltages ($V_{b2}$--$V_{b5}$). A continuous-time CMFB loop is used for simplicity, as shown in Fig.~\ref{fig:isc_opamp}(c).

The quadrature outputs ($I$ and $Q$) of the mixer are low-pass filtered by a second-order biquad~\cite{Shin_Biquad} (shown as an inset in Fig.~\ref{fig:isc_rx}) before being amplified by a differential programmable-gain amplifier (PGA). Both the biquad and the PGA use the same op-amp as in the mixer buffer. The biquad transfer function (TF) is shown in the figure. It has a Chebyshev-type response with programmable DC gain ($H(0)$), stop-band rejection ($H(\infty)$), cut-off frequency ($\omega_c$), and notch frequency ($\omega_n$). We fixed $H(0)=1$ and programmed the other biquad parameters (and thus the frequency response of the receiver) using 4-bit resistor and capacitor DACs interfaced to a standard 3-wire on-chip SPI port. The goal is to match the excitation bandwidth of $\approx f_{0}/4$ while minimizing out-of-band noise and clock feedthrough.

\subsection{Ultrasound Data Transceiver}
\begin{figure}
    \centering
    \includegraphics[width=0.78\columnwidth]{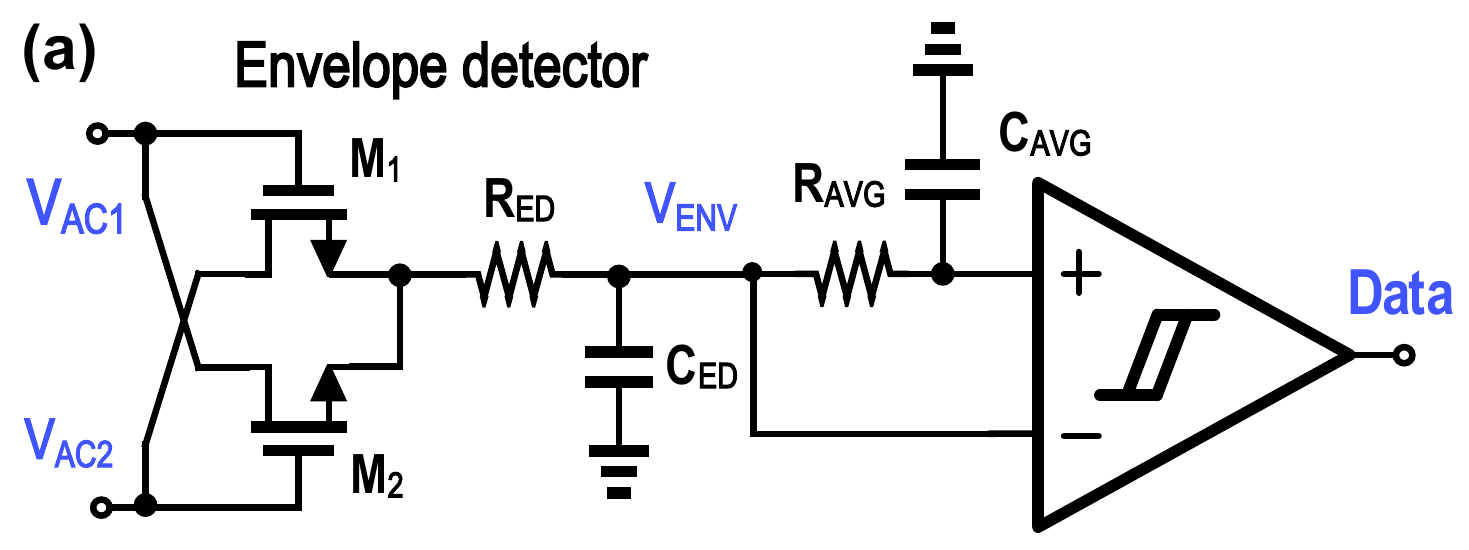}
    \includegraphics[width=0.20\columnwidth]{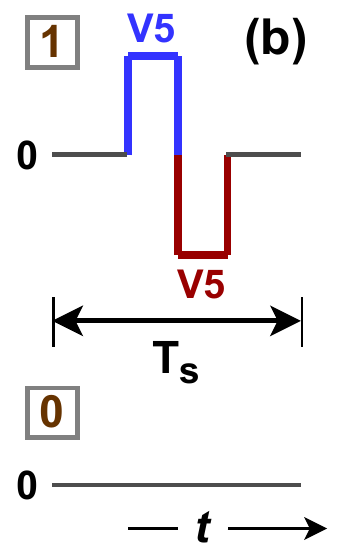}
    \caption{(a) Schematic of the receiver for the ultrasonic downlink, where $C_{ED}$ and $C_{AVG}$ are off-chip capacitors. (b) OOK symbols used for the ultrasonic uplink, where $T_{s}$ is the symbol period.}
    \label{fig:ultrasound_rx}
\end{figure}

The ultrasonic downlink and uplink reuse the two PWAS used for power delivery and SHM measurements (\#1 and \#2, respectively), as shown in Fig.~\ref{fig:isc_overview_a}(d). The hub uses binary frequency shift keying (BFSK) to modulate downlink data on the power carrier. The strongly frequency-selective ultrasound channel converts such frequency modulation into amplitude modulation, as indicated on the figure. The on-chip clock and data recovery (CDR) block uses an envelope detector (ED) and hysteretic comparator to extract the amplitude-modulated bits, as shown in Fig.~\ref{fig:ultrasound_rx}(a). Typical SHM measurement cycles are slow ($<1$ measurement/hour), so relatively low downlink data rates (10-200~bits/sec) are generally used.

Uplink data is transmitted on PWAS \#2 by reusing the SHM transmitter shown in Fig.~\ref{fig:isc_tx_rx}. For this purpose, the switches connected to voltages $V_{1}$--$V_{4}$ ($SW1$--$SW4$ and $SW6$--$SW9$) are disabled, with only $SW5$ and $SW10$ (connected to $V_{5}=3.3$~V) remaining active. Thus, the transmitter output reduces to a single-cycle pulse, as shown in Fig.~\ref{fig:ultrasound_rx}(b). Data is encoded using on-off keying (OOK) to minimize energy consumption: the presence of a pulse within a symbol period represents `1', while its absence represents `0'.

\section{Electrical Characterization Results}
\label{sec:electrical_results}

\subsection{SHM Node Design}
\begin{figure}[htbp]
	\centering
	\includegraphics[width = 0.95\columnwidth]{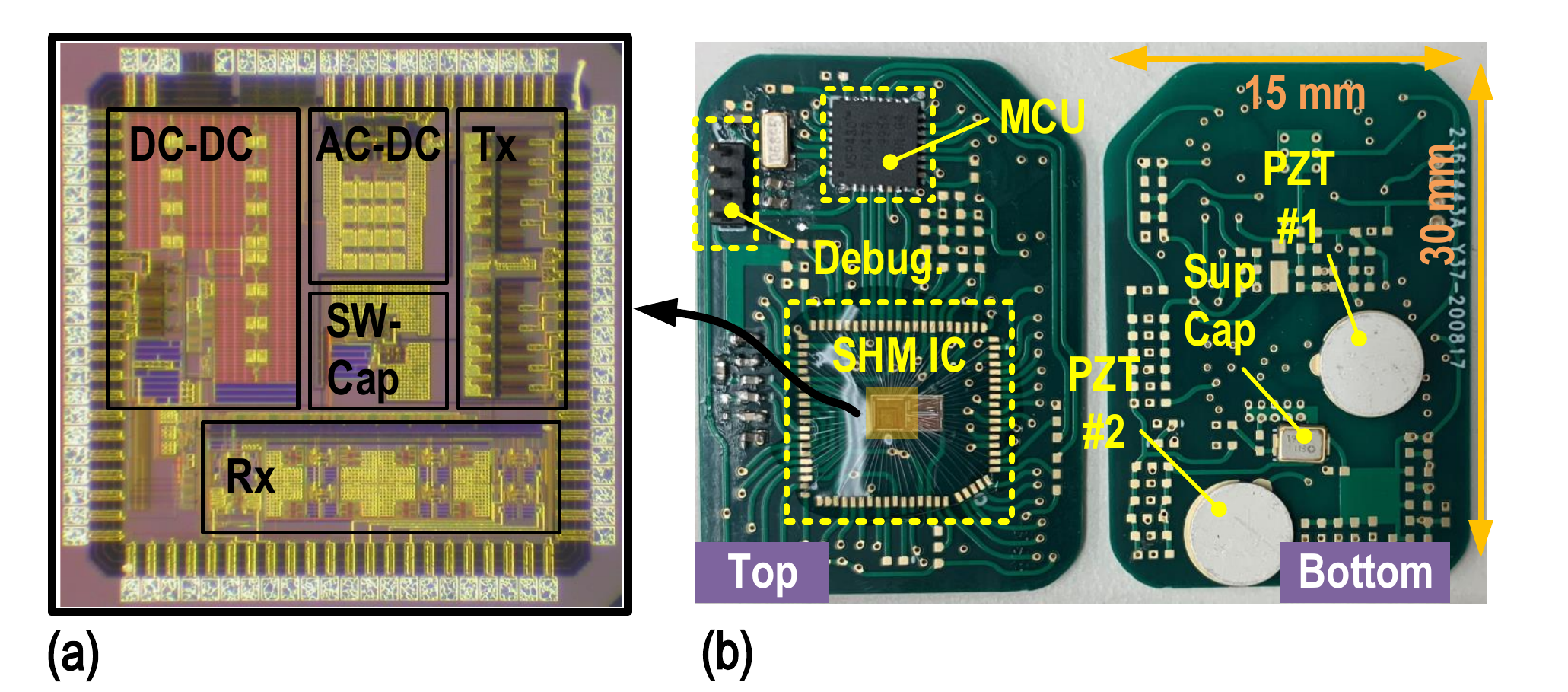}
	\caption{(a) Die photograph of the IC (size $=$ 2~mm $\times$ 2~mm); (b) photograph of the fully-autonomous wireless SHM sensor node that integrates the IC, an MCU, two PZT transducers, and passive components.}
	\label{fig:isc_die}
\end{figure}

The SHM chip was designed in the TSMC 180~nm standard CMOS process and fabricated through Muse Semiconductor. Fig.~\ref{fig:isc_die}(a) shows a labeled die photograph of the IC, which measures 2~mm $\times$ 2~mm. All on-chip bias voltages and currents were internally generated using on-chip constant-$G_m$ current references and 2T-type voltage references.

The custom SoC was integrated with an off-the-shelf ultra-low-power MCU (MSP430FR2476, Texas Instruments), two low-profile lead zirconate titanate (PZT) ultrasound transducers (diameter of 5~mm), and passive components (including $C_{RECT}=10$~$\mu$F, $C_{LOAD}=22$~$\mu$F, and $C_{STOR}=11$~mF) to realize miniaturized and fully-autonomous wireless SHM sensor nodes, as shown in Fig.~\ref{fig:isc_die}(b). Bare dies were assembled using a chip-on-board method to minimize node size (currently, 15~mm $\times$ 30~mm). This section describes electrical test results from the sensor node, while the next section presents SHM results from a test-bed.

\subsection{Active Rectifier}
\begin{figure}
    \centering
    \includegraphics[width=0.8\columnwidth]{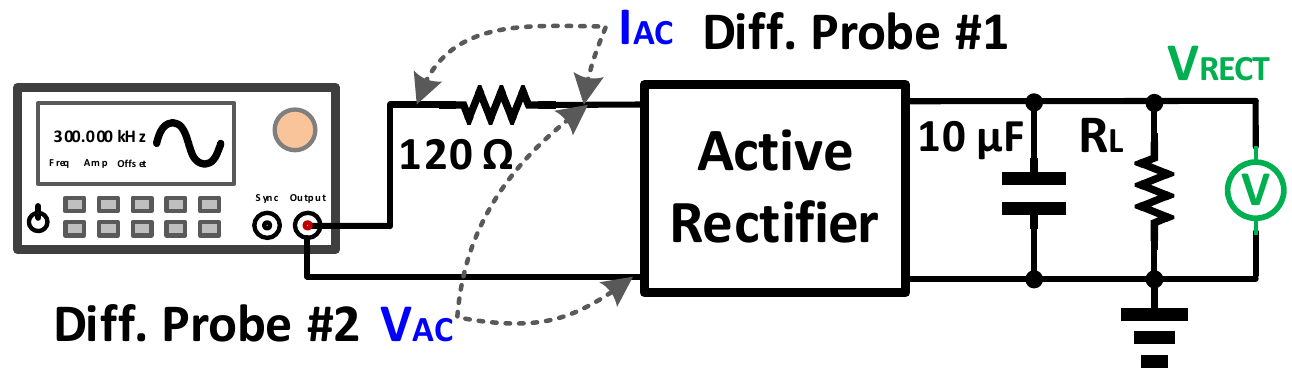}
    \caption{Experimental setup used for measuring rectifier efficiency.}
    \label{fig:efficiency_setup}
\end{figure}

Fig.~\ref{fig:efficiency_setup} shows the experimental setup used for measuring the VCE and PCE of the active rectifier. The input AC voltage and current were monitored using two differential probes, with the second connected across a 120~$\Omega$ sense resistor. The output DC voltage $V_{RECT}$ at various loads $R_L$ was monitored using a Keithley source meter unit (SMU).

\begin{figure}[htbp]
	\centering
	\includegraphics[width=0.49\columnwidth]{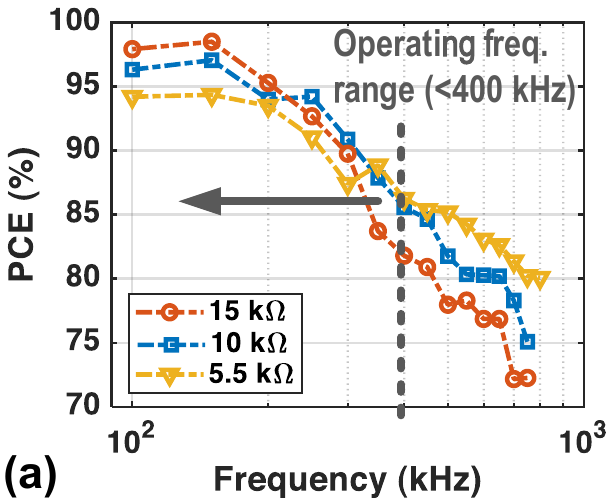}
	\includegraphics[width=0.48\columnwidth]{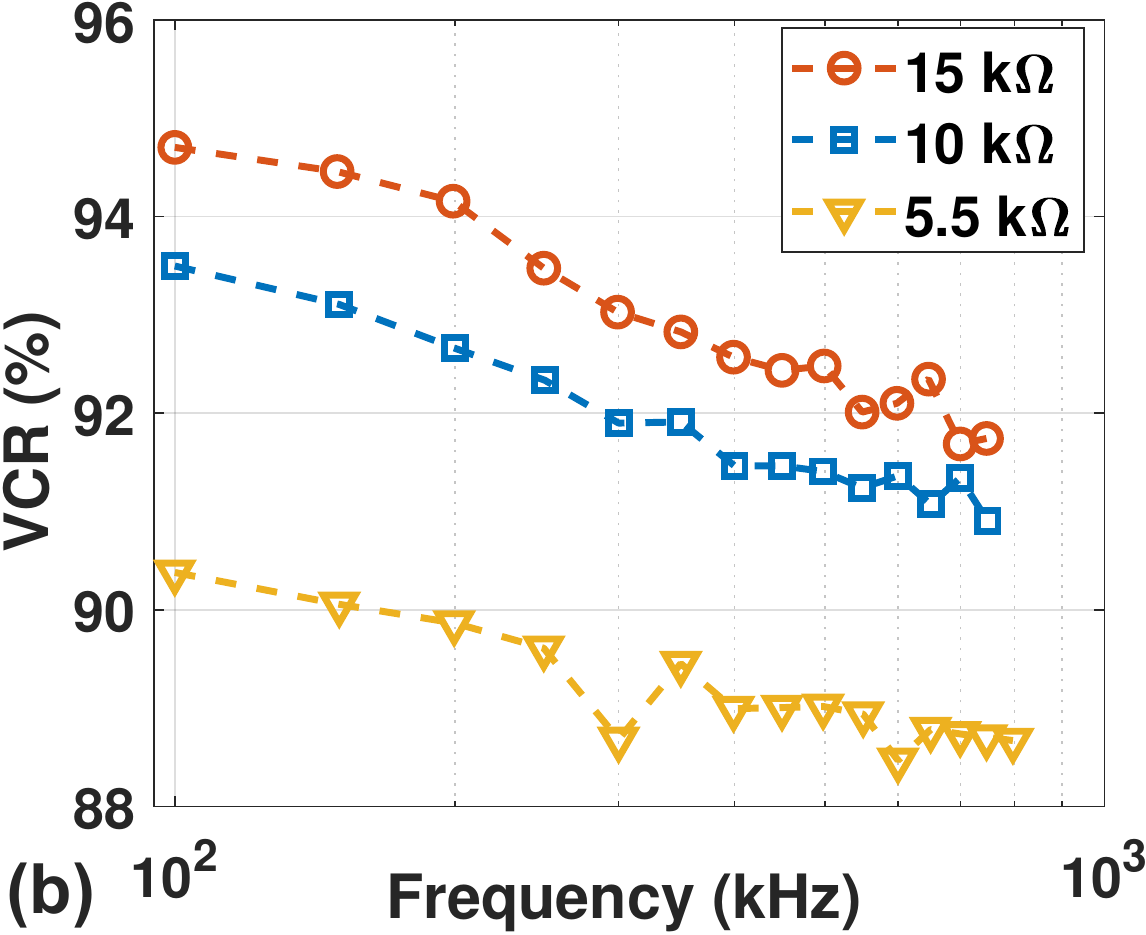}
	\caption{Measured performance of the AC-DC converter versus input frequency $f_{in}$ at different loads: (a) PCE and (b) VCR.}
	\label{fig:isc_ac_dc_results}
\end{figure}

Fig.~\ref{fig:isc_ac_dc_results} shows measured results from the proposed rectifier. Figs.~\ref{fig:isc_ac_dc_results}(a) and (b) show the PCE and VCR, respectively, as a function of input frequency $f_{in}$ for three different values of load resistance ($R_L$). Both PCE and VCR decrease with frequency due to increased dynamic loss in the switches and comparators. Also, PCE decreases with increased load (i.e., lower values of $R_L$) at low frequencies due to increased conduction loss in the switches, but the trend is reversed at high frequencies where dynamic loss (which is largely load-independent) is dominant. Excellent performance (PCE $>85$\%, VCR $>90$\%) is obtained over the expected operating range ($R_{L}=5-15$~k$\Omega$, $f_{in}<400$~kHz).

\begin{figure}[htbp]
	\centering
	\includegraphics[width=0.50\columnwidth]{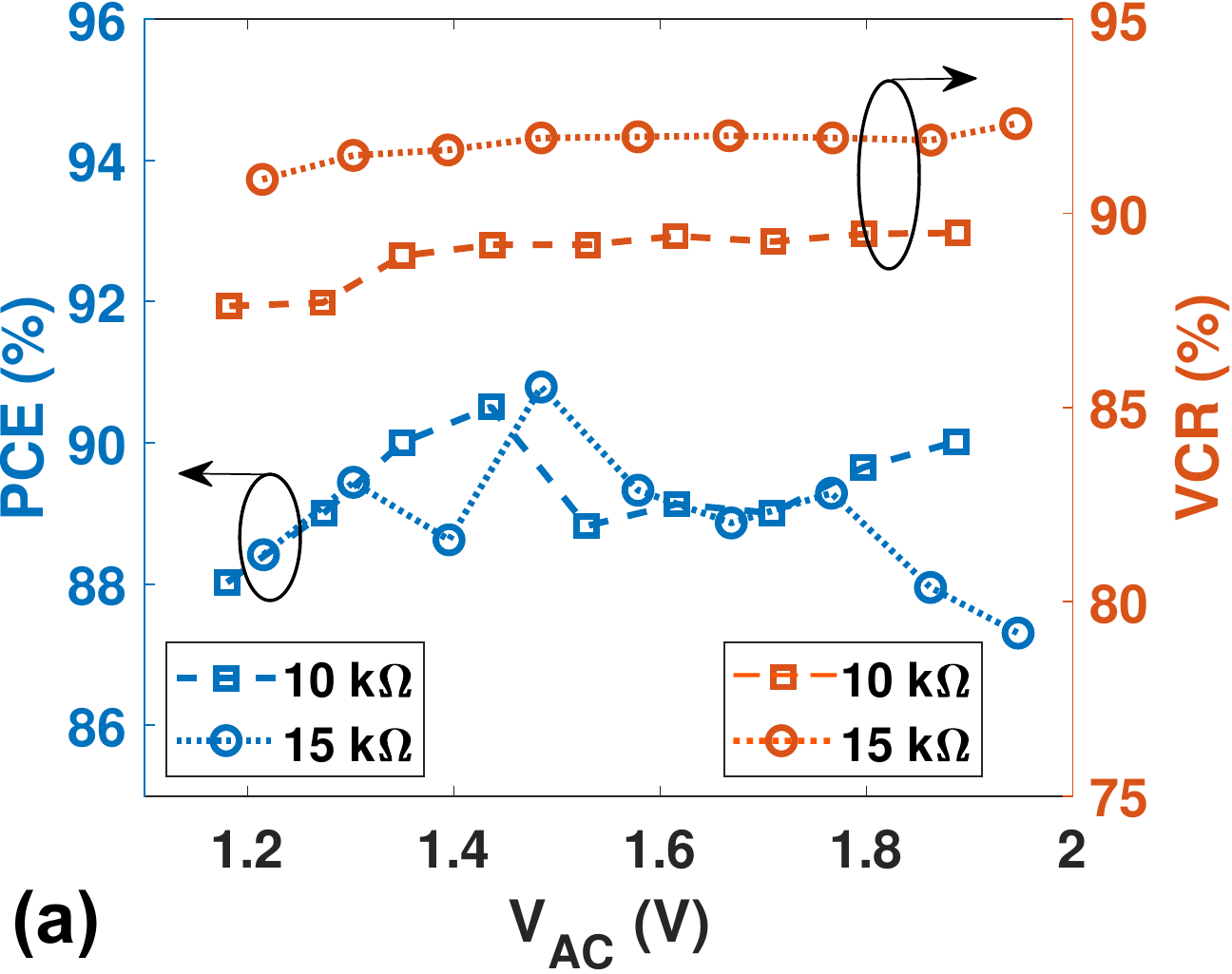}
	\includegraphics[width=0.48\columnwidth]{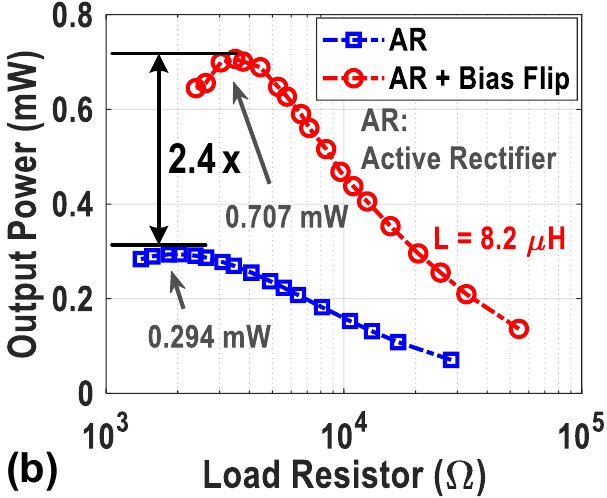}
	\caption{(a) Measured PCE and VCR of the rectifier versus AC voltage amplitude for two values of load ($R_L$). (b) Output DC power versus $R_L$ with and without the bias flip circuit. All results were obtained at $f_{in}=300$~kHz.}
	\label{fig:isc_ac_dc_biasflip}
\end{figure}

Fig.~\ref{fig:isc_ac_dc_biasflip}(a) shows that the measured PCE and VCR are nearly independent of AC input amplitude over the typical operating range ($V_{AC}=1.1-1.9$~V) at $f_{in}=300$~kHz. This result again suggests that conduction loss (which decreases with $V_{AC}$) is small compared to dynamic loss. Fig.~\ref{fig:isc_ac_dc_biasflip}(b) shows the output power as a function of $R_L$ in two cases: i) using the active rectifier (AR) alone, and ii) combining the AR with a small off-chip inductor ($L=8.2$~$\mu$H) and the bias-flip circuit. The latter increases the maximum available power by 2.4$\times$, as expected.

\subsection{DC-DC Converter}
\begin{figure}[htbp]
	\centering
	\includegraphics[width = 1\columnwidth]{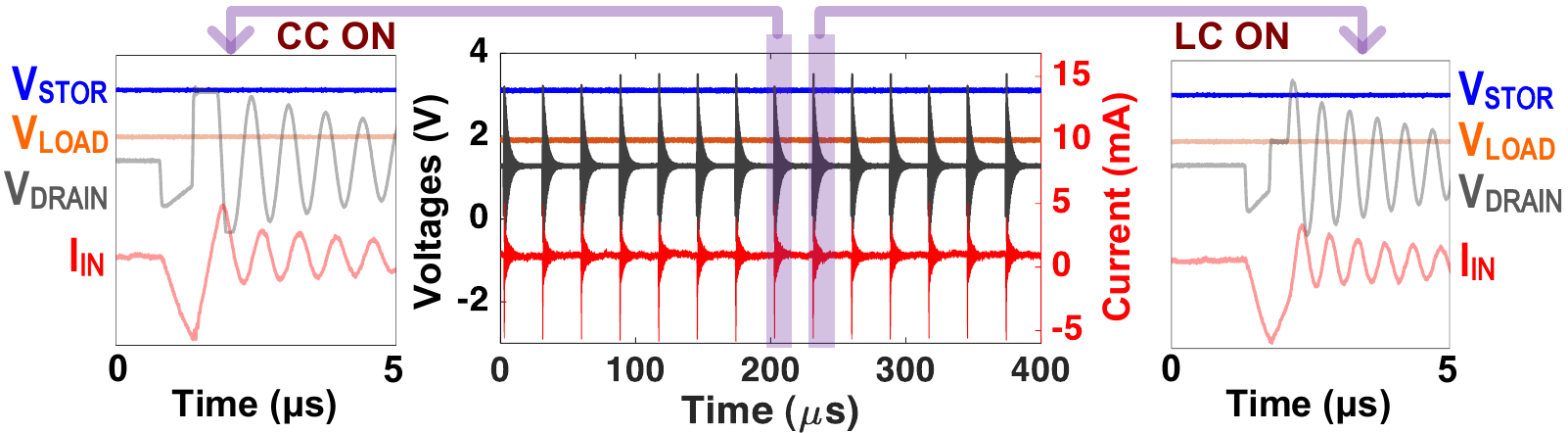}
	\caption{Typical measured waveforms for the DC-DC converter.}
	\label{fig:isc_dc_dc_meas}
\end{figure}

Typical measured waveforms for the DC-DC converter are shown in Fig.~\ref{fig:isc_dc_dc_meas}. The central figure shows a zoomed-out view, highlighting the switching frequency of $f_{s}=50$~kHz. The zoomed-in views on the left and the right show cases where the CC and the LC are ON, respectively. As expected, the voltage of the common drain node ($V_{DRAIN}$) equals $V_{STOR}$ when the CC is ON, while it equals $V_{LOAD}$ when the LC is ON. After either converter turns off, the parasitic capacitance at $V_{DRAIN}$ (which is now floating) results in high-frequency ringing. However, this does not affect the regulated voltages.

\begin{figure}[htbp]
	\centering
	\includegraphics[width = 0.48\columnwidth]{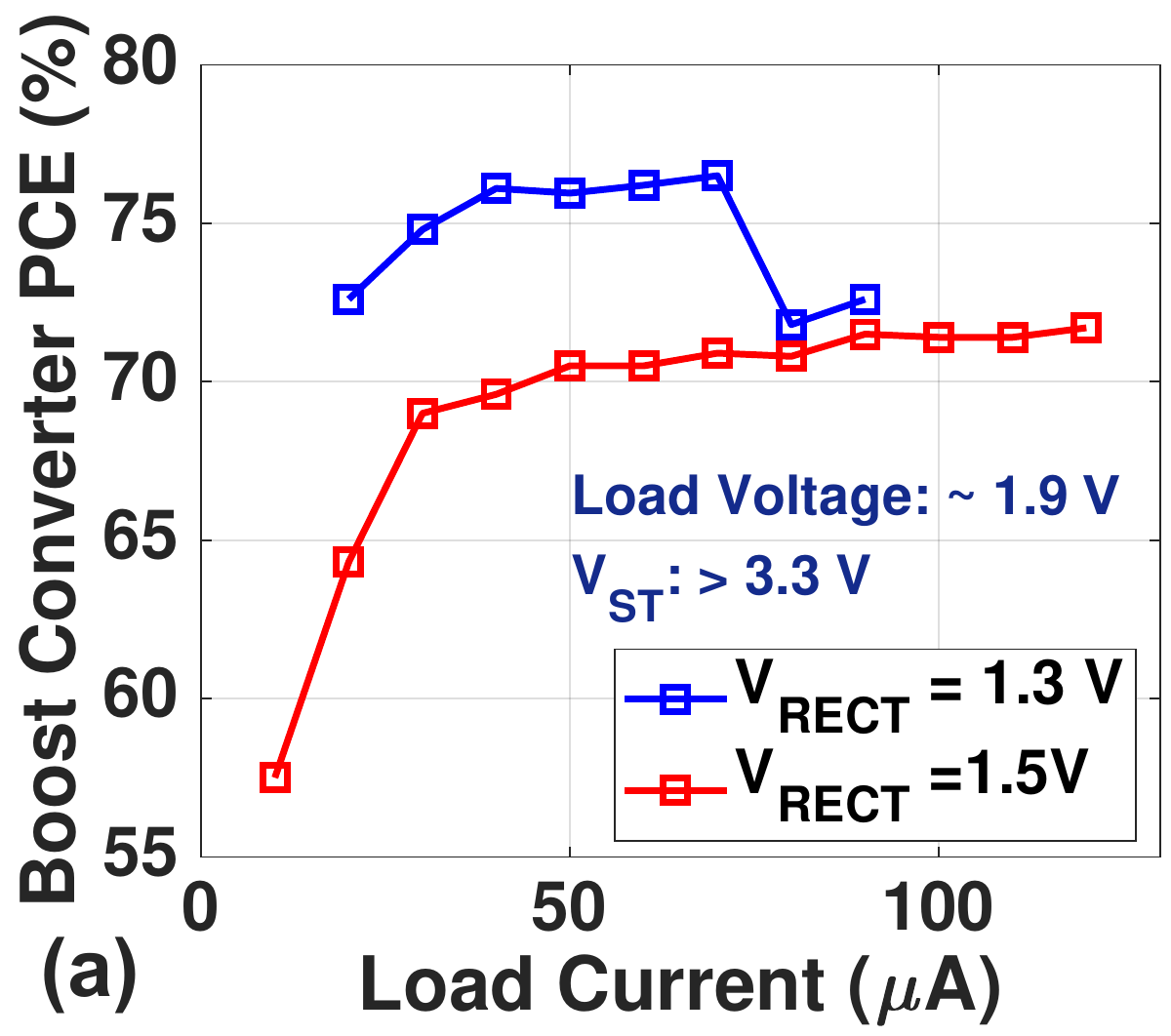}
	\includegraphics[width = 0.48\columnwidth]{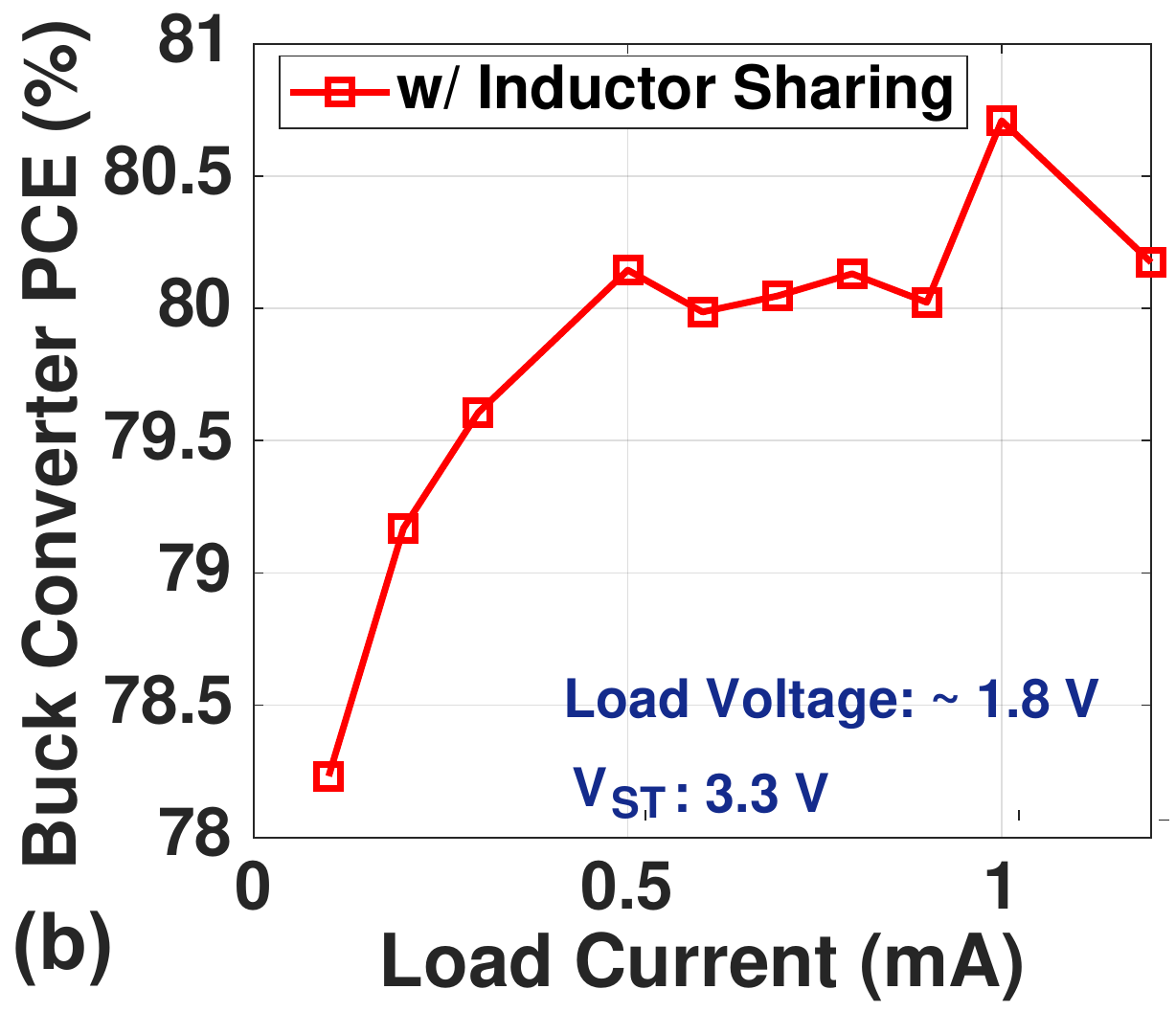}
	\caption{Measured PCE for the DC-DC converters versus load current: (a) the boost converters (CC and LC, operating simultaneously with $\alpha>0$) for different values of $V_{RECT}$; (b) the back-up buck converter (BC).}
	\label{fig:isc_dc_pce}
\end{figure}

Fig.~\ref{fig:isc_dc_pce} summarizes the measured PCE of the DC-DC converter versus load current. Fig.~\ref{fig:isc_dc_pce}(a) shows the PCE for the time-shared boost converters (CC and LC) under normal conditions, with the system cycling between states 2 and 3 such that $\alpha>0$. The figure shows that PCE $>70$\% is maintained over the typical range of $V_{RECT}$ for load currents $>30$~$\mu$A. Fig.~\ref{fig:isc_dc_pce}(b) shows the PCE for the back-up buck converter (BC), which is only active in state 1. In this case, PCE $>78$\% is maintained over a wide range of load currents.

\subsection{Switched-Capacitor DC-DC Converters}
\begin{figure}[htbp]
	\centering
	\includegraphics[width=0.52\columnwidth]{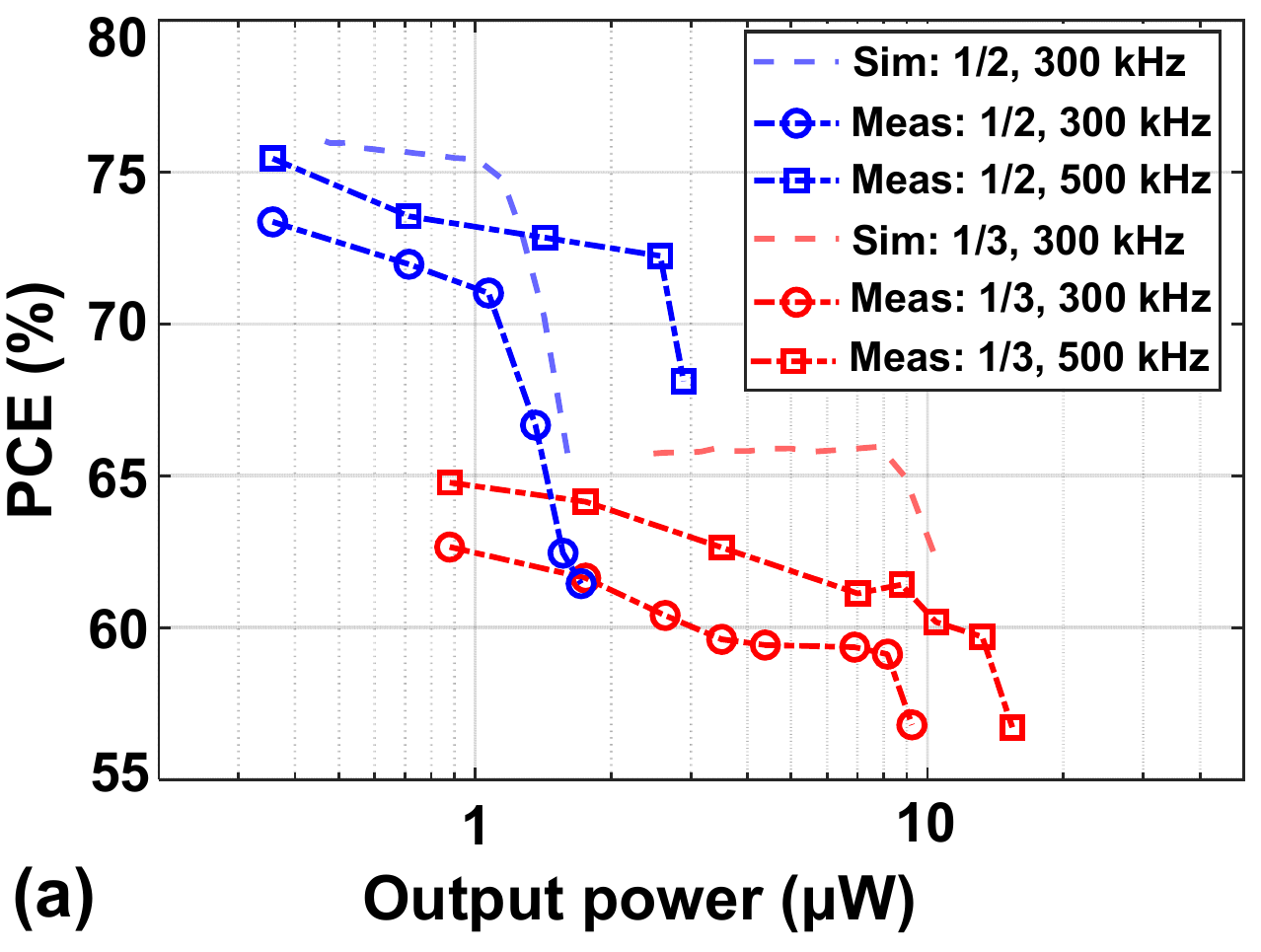}
	\includegraphics[width=0.46\columnwidth]{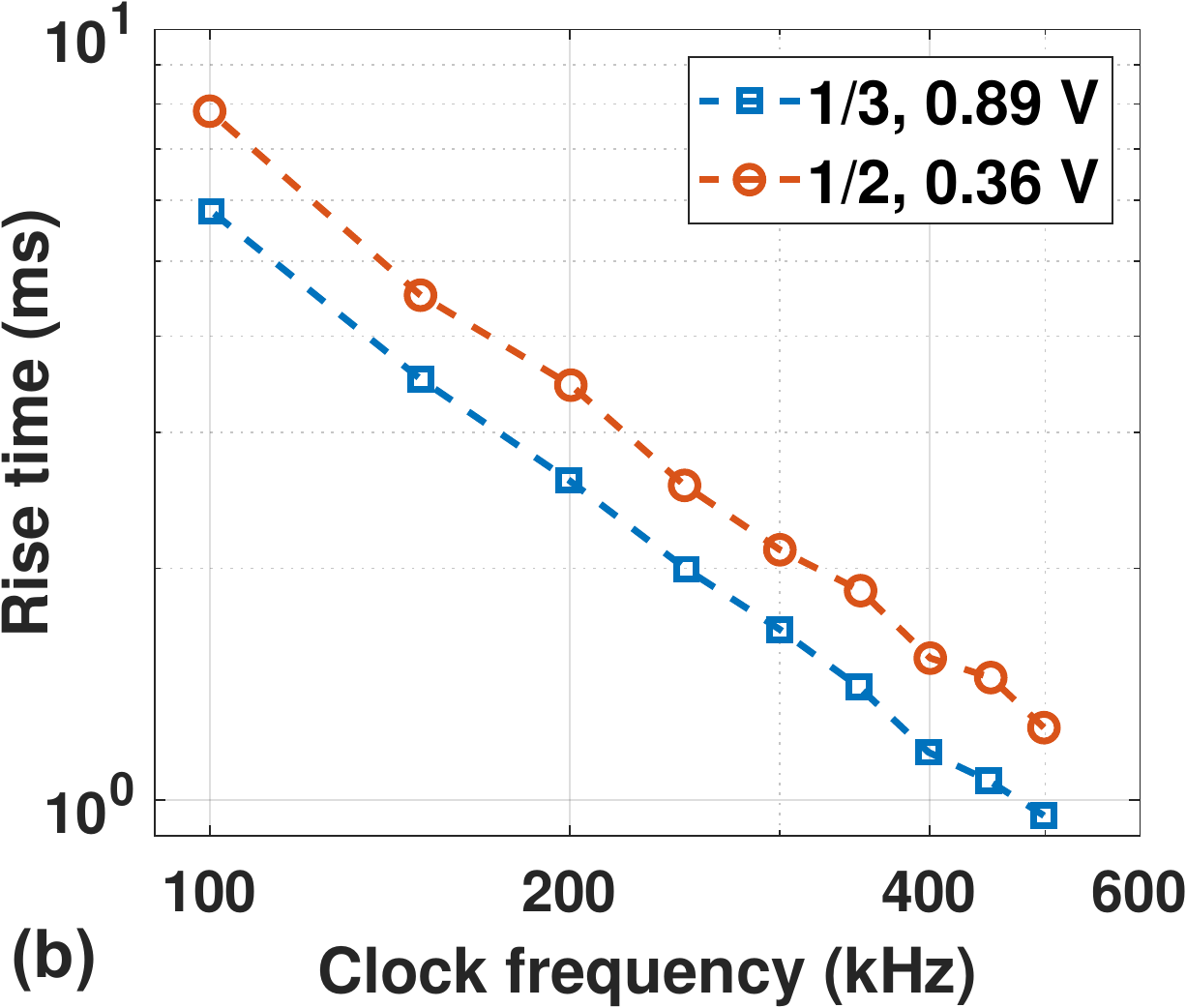}
	\caption{Measured performance of the 1/3 and 1/2 ratio switched-capacitor DC-DC converters used to power the SHM transmitter: (a) PCE versus output power for various clock frequencies and conversion ratios, compared with simulations; and (b) voltage rise time versus clock frequency.}
	\label{fig:isc_sc_results}
\end{figure}

Fig.~\ref{fig:isc_sc_results}(a) compares measured and simulated PCE of the 1/2 and 1/3 ratio switched-capacitor (SC) DC-DC converters (used by the SHM transmitter) versus output power for various conversion ratios and clock frequencies $f_{clk}$. PCE improves with $f_{clk}$, as expected; this is because the output resistance $R_{out}$ of a SC converter is a strongly-decreasing function of switching frequency. In particular, theoretical models predict $R_{out}\propto 1/f_{clk}$ when $f_{clk}$ is relatively low (known as the slow-switching limit)~\cite{seeman2008analysis}. In addition, the measured PCE is slightly higher than the simulations, probably due to higher-than-expected switching losses due to parasitic capacitance. The 1/1 converters do not contribute significantly to power loss since they use a single switch, which results in significantly higher PCE; thus, their performance is not shown here.

Fig.~\ref{fig:isc_sc_results}(b) plots the rise time of the 1/2 and 1/3 converters versus $f_{clk}$. The figure shows that rise time is inversely proportional to $f_{clk}$, which suggests that $R_{out}\propto 1/f_{clk}$ in agreement with theoretical models.

\subsection{SHM Transmitter}
\begin{figure}[htbp]
	\centering
	\includegraphics[width = 1\columnwidth]{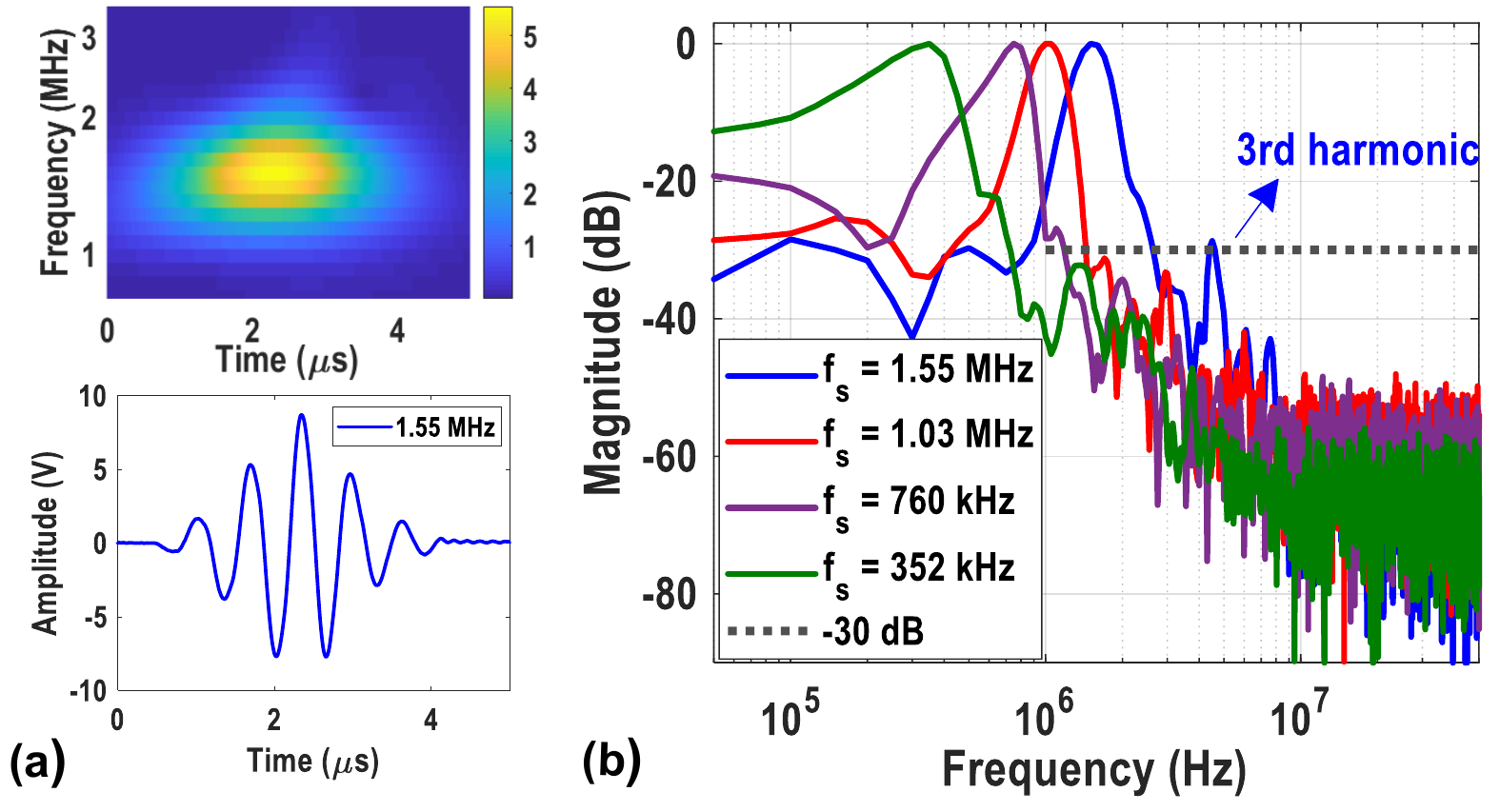}
	\caption{Measured results of the SHM transmitter: (a) typical output pulse across the transducer in the time-frequency plane (top) and time-domain (bottom); (b) output spectrum of the transmitter at various operating frequencies. The worst-case amplitude of the third harmonic is $<-30$~dBc.}
	\label{fig:isc_tx_measured}
\end{figure}

Fig.~\ref{fig:isc_tx_measured} shows electrical test results for the SHM transmitter. Fig.~\ref{fig:isc_tx_measured}(a) shows a typical output pulse across the transducer in the time-frequency plane (top) and the time-domain (bottom). The waveform closely resembles a Hamming-windowed tone burst, as desired. Note that passive amplification by the tuned $LRC$ load significantly increases the peak voltage amplitude compared to that generated by the chip ($\pm 3.3$~V). 

The resulting frequency spectra (at different operating frequencies) are shown in Fig.~\ref{fig:isc_tx_measured}(b). In each case, the off-chip series inductors ($L_1$ and $L_2$) were adjusted to match the desired operating frequency. The smooth envelope of the pulse strongly suppresses side-bands; the worst-case PSL is $\sim$30~dB, in agreement with simulations. Similarly, the tuned circuit suppresses harmonics, with worst-case out-of-band emissions (caused by the third harmonic) being smaller than $-30$~dBc.

\subsection{SHM Receiver}
\begin{figure}[htbp]
	\centering
	\includegraphics[width=0.45\columnwidth]{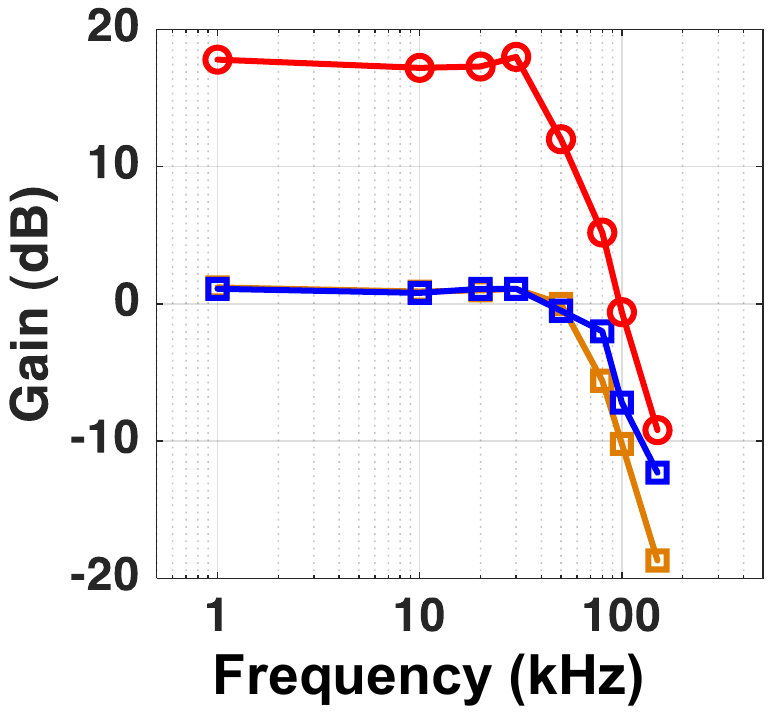}
	\caption{Measured small-signal transfer functions of the SHM receiver at different gain and bandwidth settings. These parameters can be programmed over a range of approximately 20~dB and $4\times$, respectively.}
	\label{fig:isc_rx_measured}
\end{figure}

Fig.~\ref{fig:isc_rx_measured}(a) shows measured small-signal transfer functions for the receiver at different gain and bandwidth settings; these results are in good agreement with simulations. The maximum available $-3$~dB bandwidth is $\sim$100~kHz, which is adequate over the desired range of center frequencies for SHM measurements ($f_0<400$~kHz) given that the excitation bandwidth is $\approx f_0/4$. Similarly, the maximum gain of $\sim$20~dB is adequate for SHM applications, where received signal amplitudes of 10-100~mV are typical in both transmission and reflection mode.

\subsection{Performance Summary and Comparison}
Table~\ref{tab:comparison} compares the performance of this chip with other SHM ICs in the literature, including our own earlier work~\cite{tang2018programmable,zamani2016current}. This work achieves excellent spectral localization of the transmit waveform by suppressing both sidelobes and harmonics. It is also the first to integrate ultrasonic power and data transfer capabilities within the chip, thus enabling ultrasonically-coupled SHM sensor nodes.

\begin{table*}[htbp]
    \centering
    \caption{Comparison with prior work on integrated SHM ICs.}
    \begin{tabular}{c|c|c|c|c}
    \hline
    \textbf{Parameter} & \textbf{This Work} & \textbf{CICC 2018~\cite{tang2018programmable}} & \textbf{NEWCAS 2016~\cite{zamani2016current}} & \textbf{JSSC 2014~\cite{guo2014four}} \\
    \hline
    Technology & 0.18~$\mu$m CMOS & 0.5~$\mu$m CMOS & 0.5~$\mu$m CMOS & 0.25~$\mu$m BCD \\
    Number of Tx & 1 & 1 & 1 & 4\\
    Number of Rx & 1 & 1 & 1 & --\\
    Tx Windowing Performance & $-$\textbf{36~dB (2nd harmonic)} & $-30$~dB & -- & $-40$~dB\\
    Tx Output (Maximum) & $\pm$16~V &  $\pm$6.3~V & $\pm$15~V & $\pm$36~V \\
    Tx Frequency Range & \textbf{0.05-2.5~MHz} & 0.1-2.2~MHz & 0.55-1~MHz & 780~kHz \\
    Remotely-Powered & \textbf{Yes} & No & No & No \\
    Communications & \textbf{Ultrasound half-duplex} & I$^{2}$C & No & No\\
    Tx Power Consumption & \textbf{13.2~$\mu$J~/~5~$\mu$s TX} & 875~$\mu$W & 325-575~$\mu$W & 28~$\mu$J~/~$5$~$\mu$s TX\\
    Die Size & $2\times 2$ mm$^{2}$ & $1.1\times 2.2$ mm$^{2}$ & $0.55\times 1.1$ mm$^{2}$ & $3.6\times 3.6$ mm$^{2}$\\
    \hline
    \end{tabular}
    \label{tab:comparison}
\end{table*}

\section{Active SHM measurements}
\label{sec:shm_results}
The functionality of the chip was verified by carrying out SHM measurements out on a test-bed representing an airframe panel. For this purpose, a set of wireless sensor nodes (\#1--\#6) was attached on one side of a carbon fiber reinforced polymer (CFRP) sheet (0.3~m $\times$ 0.3~m, 2~mm thick).

%

\subsection{Power Transfer}
Using the bias-flip circuit, a peak output power of $\sim$75~$\mu$W was obtained at a distance of $\sim$17~cm (from node \#3 to node \#4) for a transmit amplitude of 30~V at the optimal excitation frequency of $f_{in}=409$~kHz, as shown in Fig.~\ref{fig:SHM_power_data}(a). Similar measurements for a slightly shorter link (distance $\sim$12~cm, from node \#2 to node \#5) and closely-spaced excitation frequencies ($f_{in}=376$~kHz and 377~kHz) are shown in Fig.~\ref{fig:SHM_power_data}(b) for a transmit amplitude of 20~V. These results show that available power decreases significantly (by $\sim$25\%) when $f_{in}$ increases by only 1~kHz from its optimum value, which highlights the frequency-selective nature of the acoustic channel and the need to adaptively set $f_{in}$ for each node using nMPPT methods. It is also interesting to note that available power levels are significantly lower (by about $3\times$) than earlier experiments over similar distances using thin aluminum plates~\cite{shaik2020self}, either because of worse acoustic impedance matching with the PZT transducers or the anisotropic mechanical properties of the CFRP sheet.

\begin{figure}
    \centering
    \includegraphics[width=0.49\columnwidth]{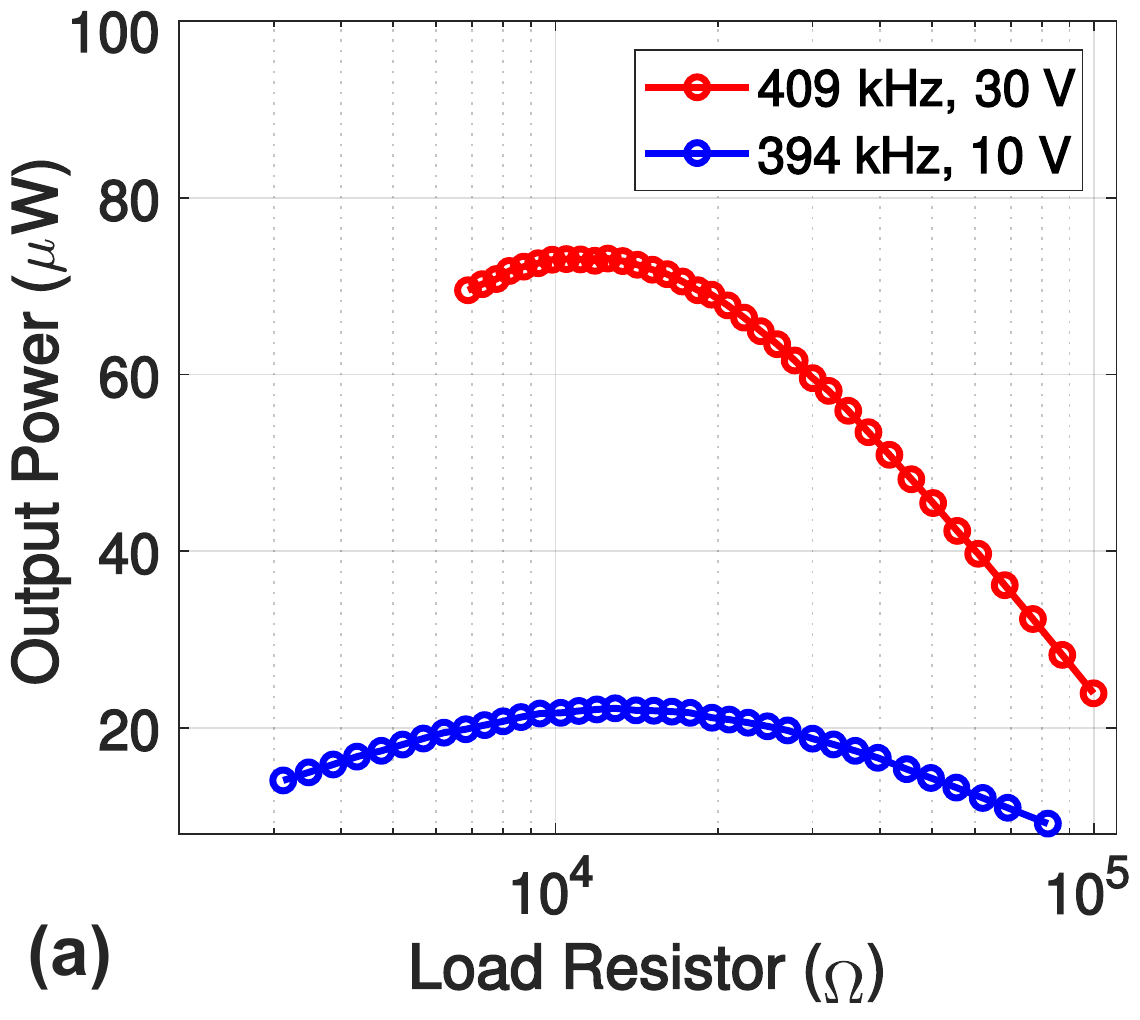}
    \includegraphics[width=0.49\columnwidth]{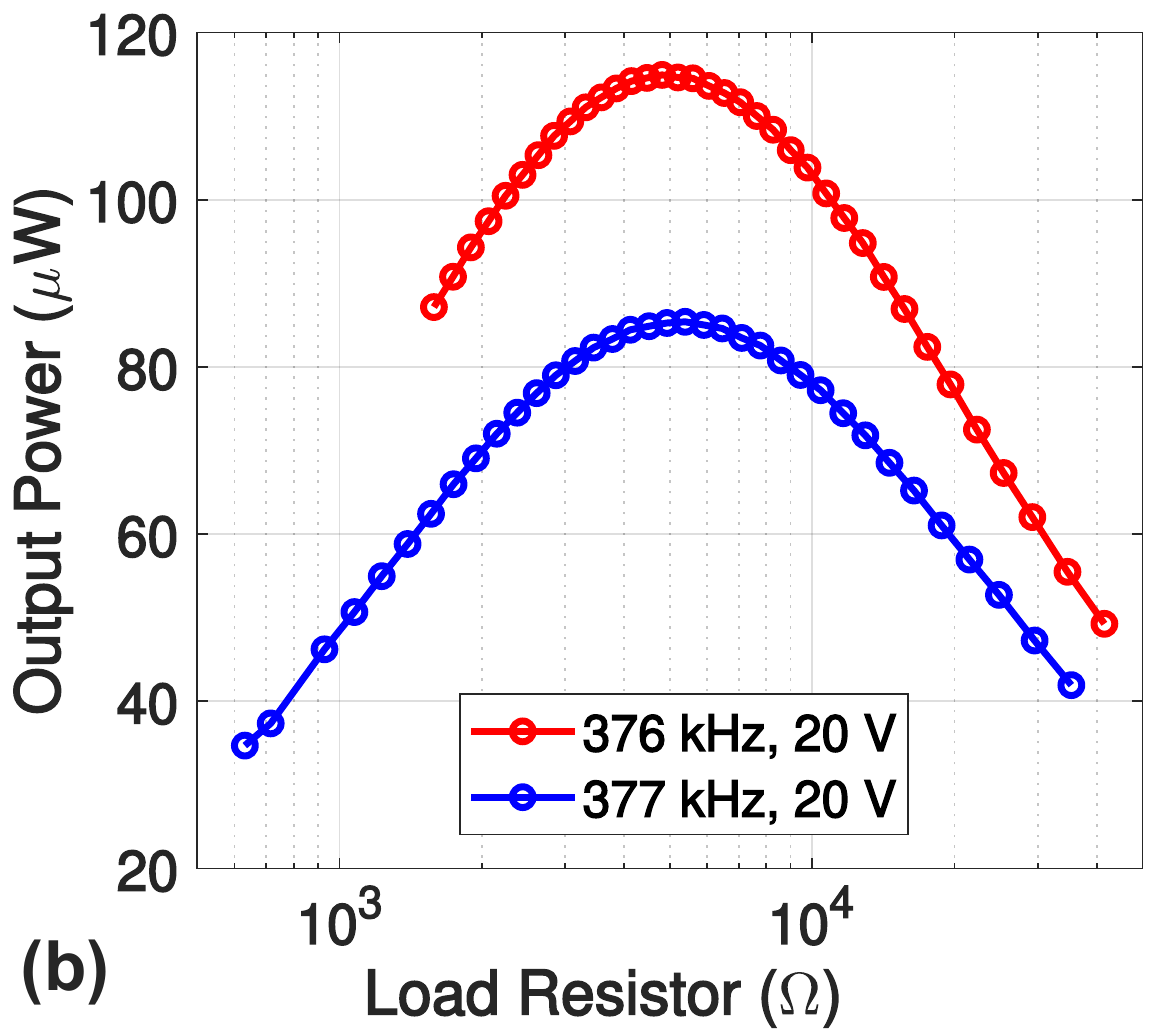}
    \caption{(a) \#3-\#4 pair power transfer; (b) \#2-\#5 pair power transfer.}
    \label{fig:SHM_pt}
\end{figure}

\subsection{Data Transfer}
\begin{figure}
    \centering
    \includegraphics[width=0.55\columnwidth]{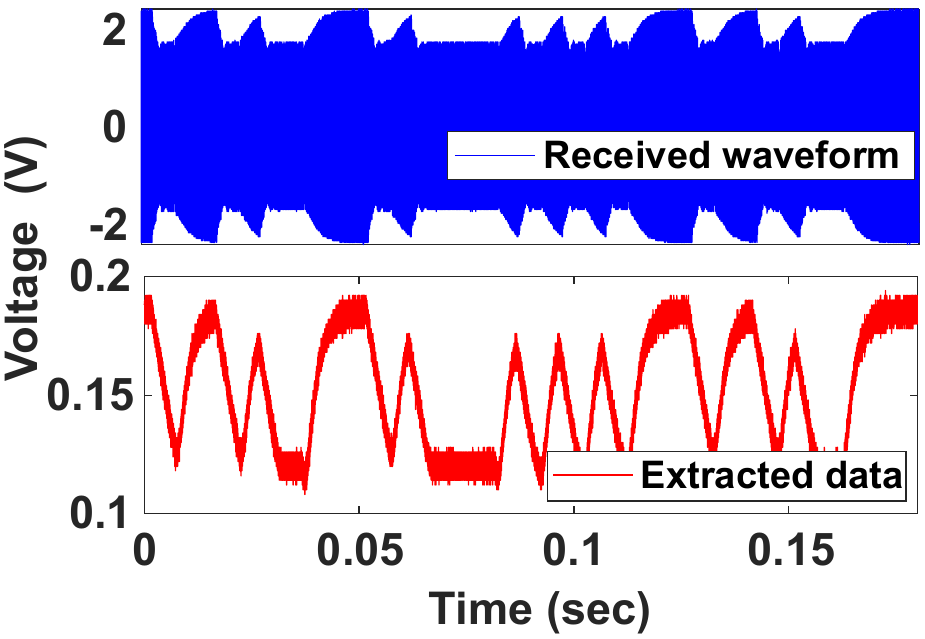}
    \caption{Typical waveforms measured for the ultrasound downlink at a bit rate of 200~bits/sec.}
    \label{fig:SHM_power_data}
\end{figure}

Fig.~\ref{fig:SHM_power_data} shows typical downlink data transmission waveforms at 200~bits/sec and a distance of $\sim$10~cm (from node \#7 to node \#3). The frequency-selective ultrasound channel converts the BFSK-modulated waveform generated by the hub (node \#7) into amplitude shift-keying (ASK), as expected. The measured bit error rate (BER) is $<10^{-4}$ (no bit errors were detected within $10^{4}$ bits). Finally, OOK-based uplink data transmission with BER $<10^{-4}$ was achieved for distances up to $\sim$20~cm and rates up to 10~kbits/s.

\subsection{Localization of Structural Damage}
A variety of damage detection and localization algorithms have been proposed for SHM using guided ultrasound waves~\cite{giurgiutiu2007structural,zhao2007active,michaels2008detection}. The general approach relies on comparing the current observation with one or more previous baselines recorded from undamaged structures, thus generating differential features that quantify changes from the baselines. These features generally include time-varying changes in the baselines due to environmental fluctuations (e.g., in temperature), which must be removed using various compensation methods~\cite{croxford2010efficient}. Finally, anomalies are detected when the compensated differential features exceed certain predefined threshold values.

\begin{figure*}
    \centering
    \includegraphics[width=0.47\columnwidth]{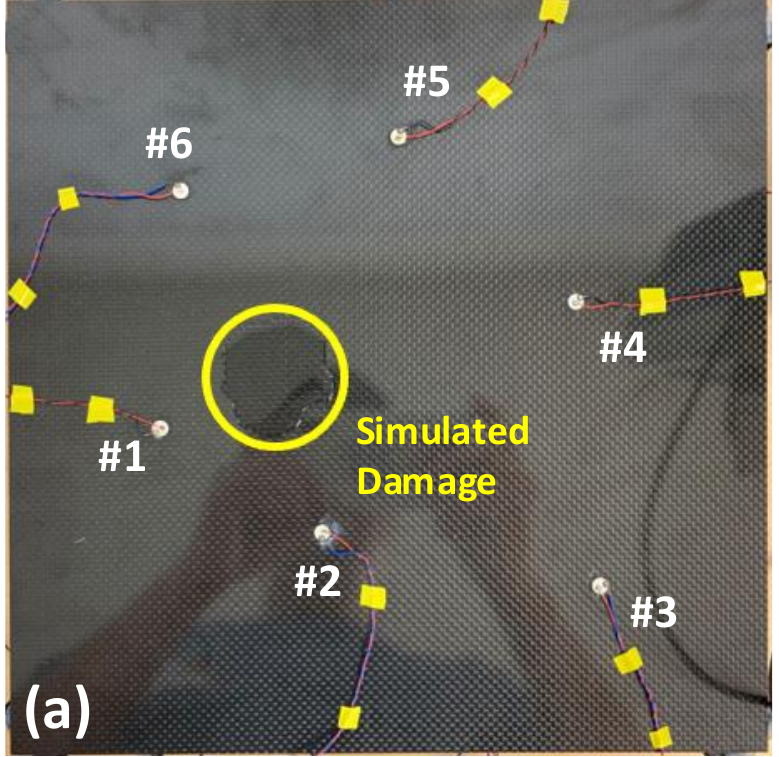}
    \includegraphics[width=0.50\columnwidth]{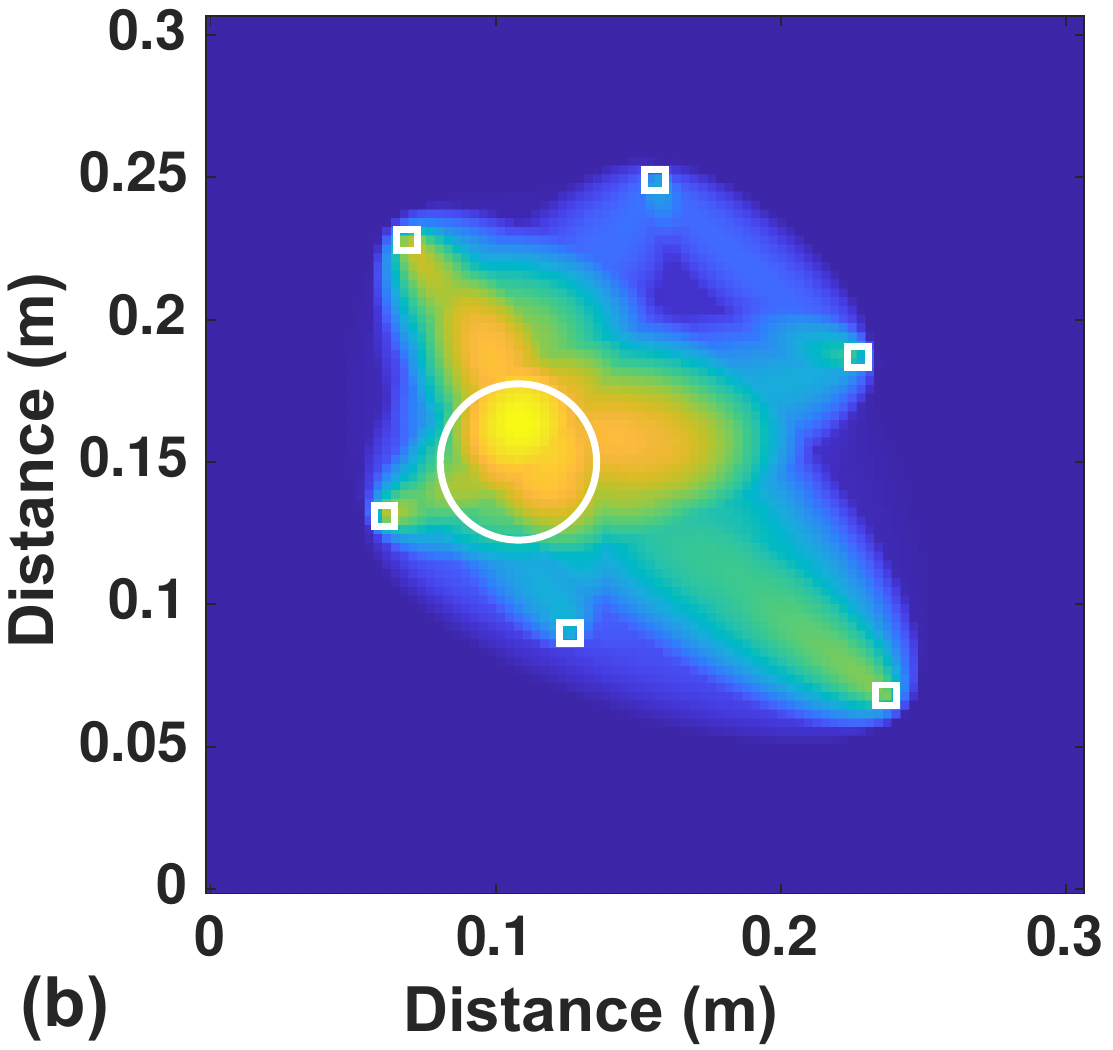}
    \includegraphics[width=0.50\columnwidth]{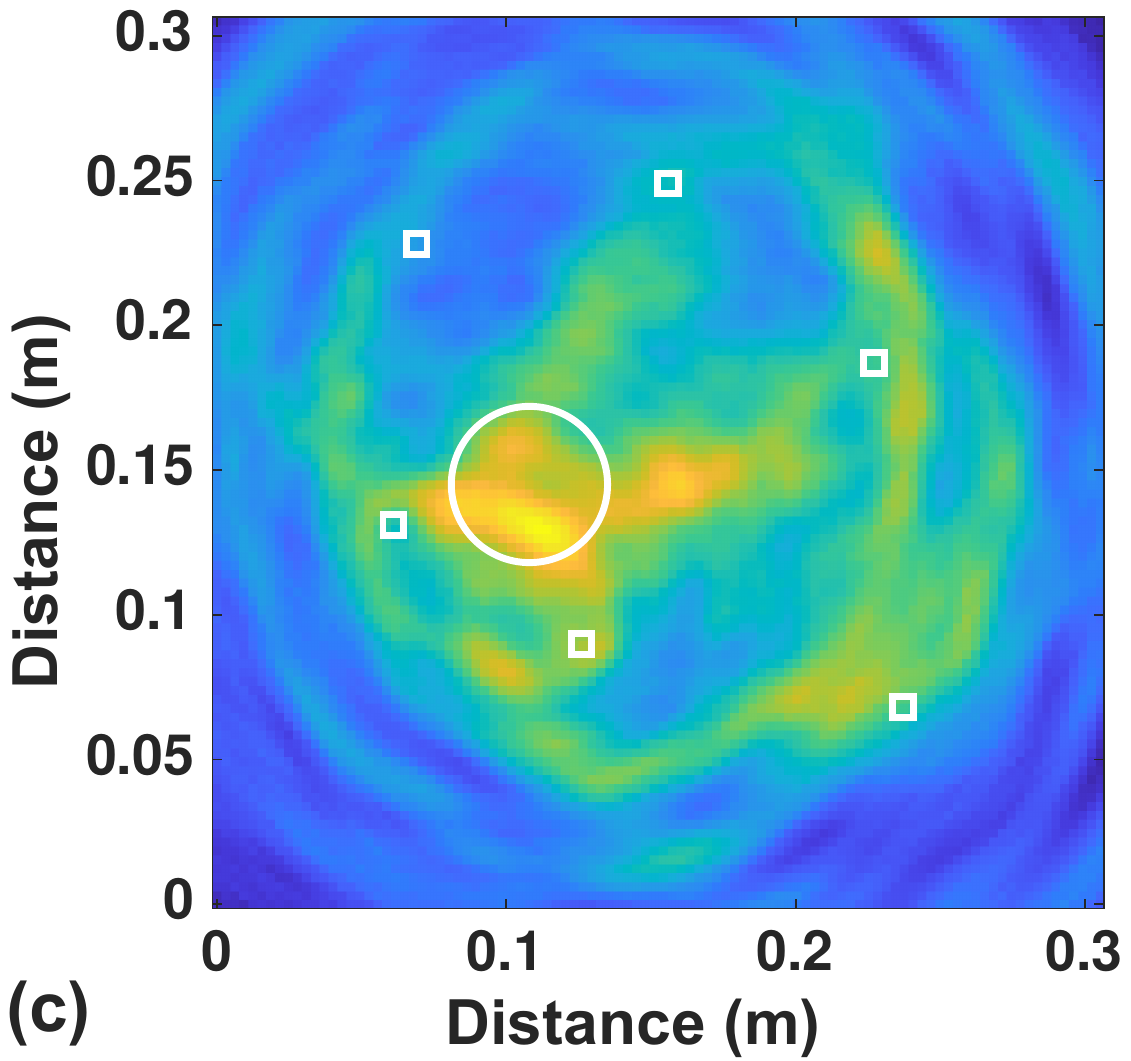}
    \includegraphics[width=0.50\columnwidth]{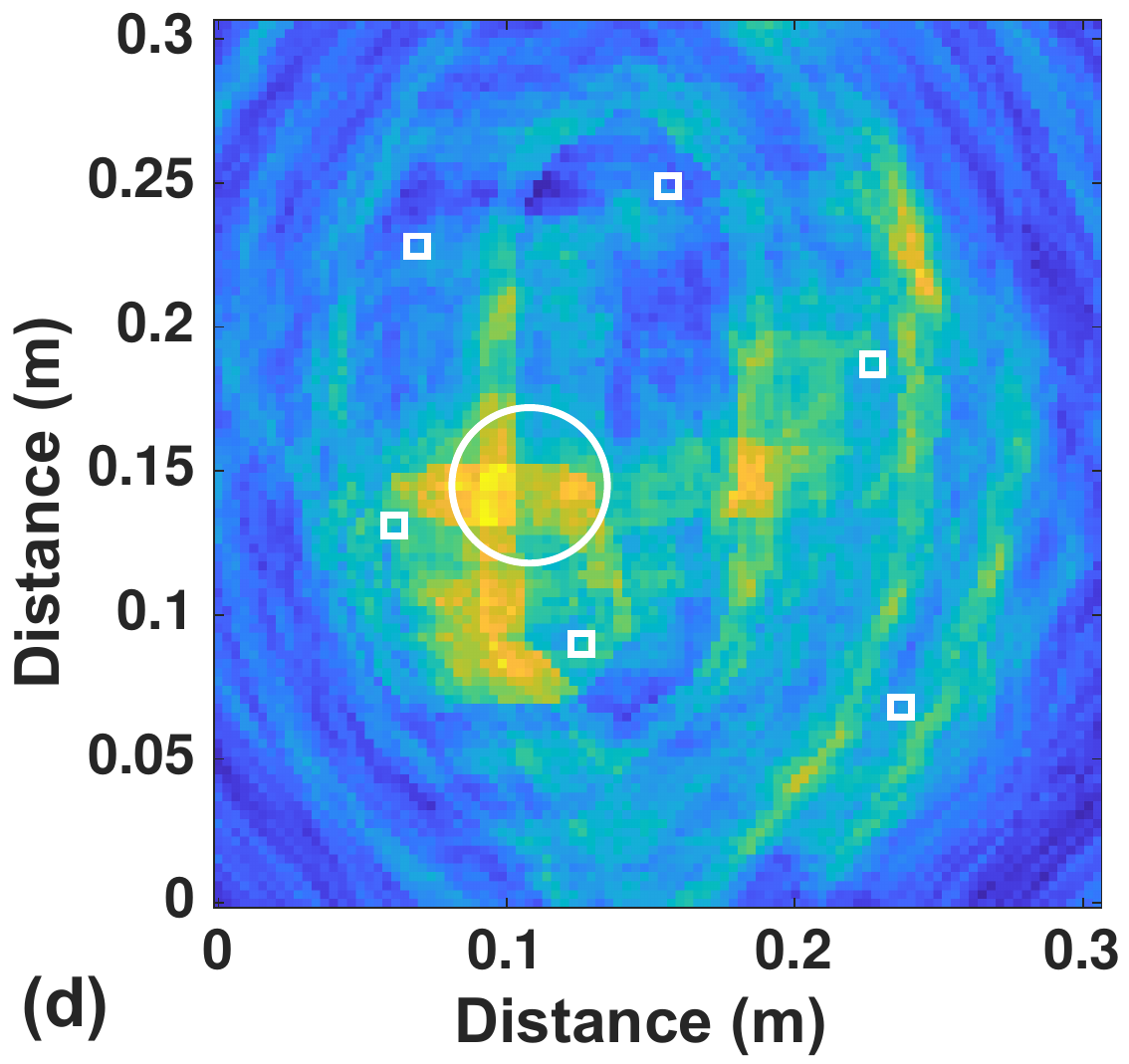}
    \includegraphics[width=0.065\columnwidth]{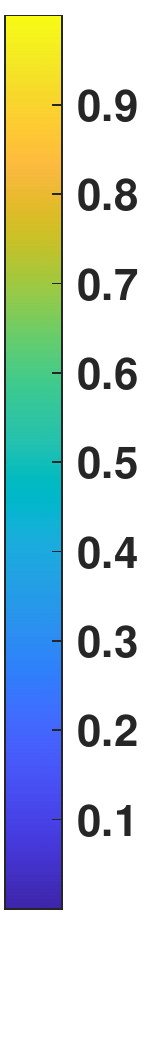}
    \caption{(a) Photograph of the SHM test-bed, which uses six wireless sensor nodes placed on on side of a 2-mm-thick CFRP sheet (size $=0.3\times 0.3$~m). Structural damage was simulated by placing a drop of water on the sheet (as indicated by the yellow circle). (b)-(d) Damage localization maps obtained using: (b) RAPID, (c) delay-and-sum (DAS), and (d) group velocity compensated DAS algorithms, respectively.}
    \label{fig:SHM_testbed}
\end{figure*}

A set of 6 wireless sensor nodes were used for initial damage localization experiments on the SHM test-bed. For this purpose, a drop of water was placed on the surface of the CFRP sheet to locally perturb the Lamb wave velocity and thus simulate structural damage, as shown in Fig.~\ref{fig:SHM_testbed}(a). Data for a single measurement was acquired by transmitting from one node and receiving from all the nodes. The process was then repeated with each of the 6 sensor nodes serving as the transmitter, thus resulting in a $6\times 6$ data matrix. 

Data matrices obtained with and without the simulated damage were then processed off-line using two well-known SHM algorithms, namely RAPID (Reconstruction Algorithm for Probabilistic Inspection of Damage)~\cite{zhao2007active} and delay-and-sum (DAS)~\cite{tang2018sparse}, to extract damage localization maps. The resulting maps, which are shown in Figs.~\ref{fig:SHM_testbed}(b)-(c), encode the probability that damage is present at each spatial point $(x,y)$ on the test-bed. The results show that both RAPID and DAS algorithms can successfully localize the simulated damage. However, the RAPID algorithm relies on combining pairwise measurements and thus has poor sensitivity to points outside the convex hull of sensor node positions, which reduces its usefulness for large-area SHM. While the DAS algorithm does not suffer from this problem, a close examination of the resulting damage map (Fig.~\ref{fig:SHM_testbed}(c)) shows relatively poor spatial resolution and several unwanted maxima outside the damage region. These issues can be addressed by using a more accurate Lamb wave propagation model. Due to the anisotropic mechanical properties of the CFRP sheet, the group velocity acquires a non-uniform angular dependence $v_{g}(\theta)$ that affects the DAS results. This dependence was experimentally characterized for the test-bed by using pair-wise propagation measurements and then included in the DAS algorithm. The resulting group velocity compensated damage map is shown in Fig.~\ref{fig:SHM_testbed}(d); it has significantly improved spatial resolution, as expected.

\section{Conclusion}
\label{sec:conclusion}
This paper has described a highly-integrated SoC that enables ultrasonically-coupled wireless SHM networks on structures. Electrical test results confirm the functionality of all major on-chip blocks, including the PMU, SHM transceiver, and ultrasound data transceiver. The chip was used to realize autonomous sensor nodes that were successfully deployed on an SHM test-bed (CFRP panel). Future work will focus on further miniaturization of the sensor nodes by integrating a two-channel ADC, clock generator, FSM, and SRAM on the chip, thus eliminating the external MCU.

\section*{Acknowledgment}
The authors would like to thank Prof. Joel Harley for insightful discussions, and Mohammed Sameer Shaik for assistance with the experiments.

\ifCLASSOPTIONcaptionsoff
  \newpage
\fi


\bibliographystyle{IEEEtran}
\bibliography{IEEEabrv,references}

\EOD

\end{document}